\documentclass[aps,prl,preprint,superscriptaddress,longbibliography ]{revtex4-1}

\usepackage{graphicx}
\usepackage{placeins}
\usepackage{etoolbox} 
\usepackage{mathrsfs}
\usepackage{amsfonts}
\usepackage{amsmath}
\usepackage{amssymb}
\usepackage{bm}
\usepackage{setspace}

\usepackage[version=4]{mhchem}
\usepackage{physics}
\usepackage[separate-uncertainty = true,multi-part-units=single]{siunitx}
\usepackage{natbib}

\usepackage{xr-hyper}
\usepackage[backref=none,bookmarksnumbered=true,bookmarks=true,bookmarksopen=true,colorlinks=true,citecolor=blue,linkcolor=blue,filecolor=blue,anchorcolor=green,urlcolor=blue,unicode=false]{hyperref}

\usepackage{ulem}[normalem] 
\def\bb{2B}
\def\bgnr{\bb-GNR}

\def\z2b{$d_{\scriptsize\mbox{{2B}}}$}

\newcommand*\didv{\mathrm{d} I/\mathrm{d} V}

\makeatletter
\patchcmd{\acs@contact@details}{E}{*\,E}{}{}
\makeatother

\begin{document}

\title{Tunable Current Rectification Through a Designer Graphene Nanoribbon} 
  \author{Niklas Friedrich} \email{n.friedrich@nanogune.eu}
         \affiliation{CIC nanoGUNE-BRTA, 20018 Donostia-San Sebasti\'an, Spain}

  \author{Jingcheng Li}
         \affiliation{CIC nanoGUNE-BRTA, 20018 Donostia-San Sebasti\'an, Spain}
         \affiliation{School of Physics, Sun Yat-sen University, Guangzhou 510275, China}

  \author{Iago Pozo}
        \affiliation{CiQUS, Centro Singular de Investigaci\'on en Qu\'{\i}mica Biol\'oxica e Materiais Moleculares, 15705 Santiago de Compostela, Spain}
        
   \author{Diego Pe\~na}  
        \affiliation{CiQUS, Centro Singular de Investigaci\'on en Qu\'{\i}mica Biol\'oxica e Materiais Moleculares, 15705 Santiago de Compostela, Spain}
  
  \author{Jos\'e Ignacio Pascual} \email{ji.pascual@nanogune.eu}
        \affiliation{CIC nanoGUNE-BRTA, 20018 Donostia-San Sebasti\'an, Spain}
        \affiliation{Ikerbasque, Basque Foundation for Science, 48013 Bilbao, Spain}

\date{\today}

\begin{abstract} \setstretch{1.2} 
\textbf{Unimolecular current rectifiers are fundamental building blocks in organic electronics.  Rectifying behavior has been identified in numerous organic systems due to electron-hole asymmetries of orbital levels interfaced by a metal electrode. As a consequence, the rectifying ratio (RR) determining the diode efficiency remains fixed for a chosen molecule-metal interface.     
Here, we present a mechanically tunable molecular diode exhibiting an exceptionally large rectification ratio ($> 10^5$) and reversible direction. The molecular system comprises a 7-armchair graphene nanoribbon (GNR) doped with a single unit of substitutional diboron within its structure, synthesized with atomic precision on a gold substrate by on-surface synthesis. The diboron unit creates half-populated in-gap bound states and splits the GNR frontier bands into two segments, localizing the bound state in a double barrier configuration. By suspending these GNRs freely between the tip of a low-temperature scanning tunneling microscope and the substrate, we demonstrate unipolar hole transport through the boron in-gap state's resonance. Strong current rectification is observed, associated with the varying widths of the two barriers, which can be tuned by altering the distance between tip and substrate. This study introduces an innovative approach for the precise manipulation of molecular electronic functionalities, opening new avenues for advanced applications in organic electronics.
}\\
\end{abstract}

\maketitle

\section{Introduction}
Quantum electron transport through single molecules offers the possibility of replicating functionalities of electronic devices with customized organic elements. Such a basic idea is the foundation of organic electronics, as launched by the seminal proposition of Aviram and Ratner for a molecular structure behaving as a diode \citep{Aviram1974}. Systematic measurements found that non-linearities in electronic conductance are ubiquitous in molecular systems \citep{gupta_nanoscale_2023}, owing to their discrete orbital level alignment \citep{yamada_electrical_2008, gehring_single-molecule_2019}. Furthermore, the flexible character of molecules offers novel schemes of current rectification, \textit{e.g.}, related to induced conformational changes \citep{Chen2017, atesci_humidity-controlled_2018}. To date, single-molecule-based rectifiers are mostly small soluble molecules \citep{gupta_nanoscale_2023} and offer limited \textit{in situ} control over key characteristics like their current rectification ratio (RR) \citep{batra_tuning_2013, xin_stereoelectronic_2017, nguyen_control_2017} or rectification direction \citep{yuan_controlling_2015}.

In recent years, the combination of in-solution synthesis of organic precursor molecules with thermally-driven on surface synthesis strategies (OSS) \citep{Clair2019} successfully integrated different functionalities in single graphene nanoribbons (GNRs). Designer GNR architectures were synthesized incorporating sharp donor-acceptor interfaces \citep{cai_graphene_2014, chen_molecular_2015}, quantum dots \citep{Carbonell-Sanroma2017, rizzo_rationally_2021}, magnetic units \citep{Li2018, friedrich_addressing_2022}, or even luminescence active elements \citep{senkovskiy_making_2017, Chong2018}. In these complex systems, sharp potential barriers were precisely inserted through the shape \citep{chen_molecular_2015, rizzo_rationally_2021}, the atomic substitution of heteroatoms \citep{cai_graphene_2014, friedrich_magnetism_2020} or by attaching functional groups \citep{Carbonell-Sanroma2017_edge}, and their functionality was corroborated by precision spectroscopic measurements utilizing scanning tunneling microscopy.  
While lithography strategies are being developed to towards the integration of graphene nanoribbons into three-terminal devices \cite{pei_exchange-induced_2022, niu_exceptionally_2023, chen_phase-coherent_2023, Zhang2023}, highly controlled two-terminal experiments on GNRs suspended between the tip and substrate have already revealed with precision transport modes through individual GNRs \citep{Lafferentz2009, Koch2012, Jacobse2018, Li2019, lawrence_probing_2020, friedrich_magnetism_2020, friedrich_addressing_2022, jiang_length-independent_2022}.

In this article, we demonstrate the extraordinary current rectification of an atomically precise engineered GNR diode exhibiting a large mechanically tunable RR ($>10^5$), which can be controlled over several orders of magnitude and also be bias-reversed. The GNR consists of a narrow armchair graphene nanoribbons (a seven atom wide armchair GNR, 7AGNR) containing a single unit of substitutional diboron dopant inside (\bgnr s). The boron doped GNR is synthesized on a Au(111) surface by diluting a small amount of borylated bisbromo-trianthracene organic molecules with the bianthracene precursors normally used to form 7AGNRs \cite{Carbonell-Sanroma2017,friedrich_magnetism_2020}. The diboron unit induces two singly-occupied states of topological origin \citep{friedrich_magnetism_2020} localised at either side of the dopant. The two states mix forming a doubly-degenerate in-gap bound state extending to both sides of the \bb-dopant (Fig.~\ref{fig:Fig1}\textbf{a}) \cite{friedrich_magnetism_2020, Zhang2022}. This boron-induced in-gap state is half-occupied with two electrons. Consequently, it appears split into the occupied (O2B) and unoccupied (U2B) spectral levels shown in Fig.~\ref{fig:Fig1}\textbf{b}, which enable hole and electron resonances for the electrical transport, respectively. Importantly, the \bb-dopant is also a potential barrier for the frontier bands of the 7AGNR \cite{Carbonell-Sanroma2017}, which are then split in two segments and laterally confined to form quantum well states (VB$_{n\text{T}}$ and VB$_{n\text{S}}$, $n=1, 2, 3, ...$ in Fig.~\ref{fig:Fig1}\textbf{b}). In our experiments, we investigated quantum transport through a \bgnr\ suspended between the tip and the surface of an STM and found fingerprints of unipolar hole transport at both bias polarities.
Our results reveal a mechanism for asymmetric hole transmission governed by the relative lengths of the two 7AGNR segments, defined by the tip-substrate distance. This mechanism forms the basis of the observed tunable current rectification.

\section{Unipolar Transport Through the \bb-state}

\begin{figure*}[t]
	\includegraphics[width=\textwidth]{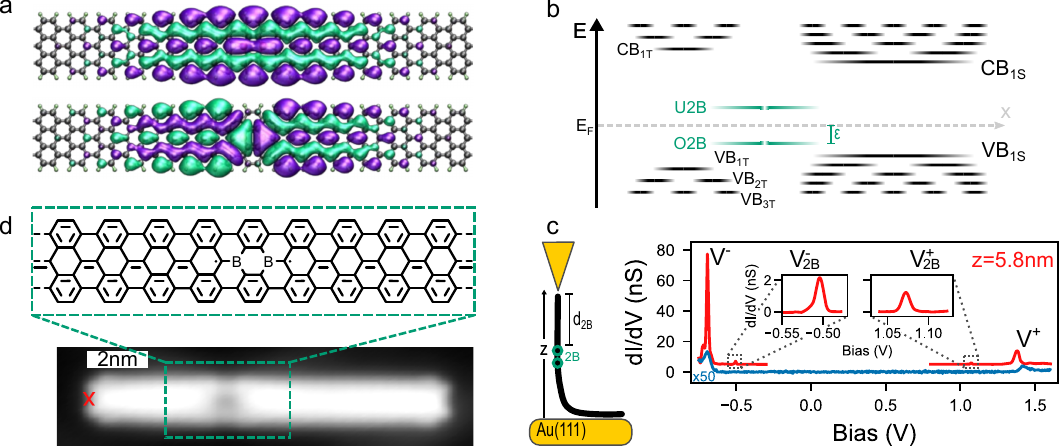}
	\caption{\label{fig:Fig1} 
	\textbf{Bandstructure of a free-standing \bb-GNR:} 
	\textbf{a} Wavefunctions of the degenerate O2B \citep{friedrich_magnetism_2020} originating from the symmetric and anti-symmetric linear combination of the topological boundary states. The spins in the two O2B orbitals are aligned.
    \textbf{b} Sketch of the ribbon's band structure in real space. The VB and CB are confined in two separate segments by the \bb-unit resulting in quantum well states. The segment's different length results in different band onset energies. The topological boundary state emerges around the \bb-unit inside the band gap.
	\textbf{c} $\didv$ spectra for a \bb-GNR (red, $z=\SI{5.8}{nm}$) and a pristine 7AGNR (blue, $z=\SI{6}{nm}$). The two dominant features $V^-$ and $V^+$ are present in both systems and attributed to resonant band transport, while the two peaks $V_\text{2B}^-$ and $V_\text{2B}^+$ appear between the band transport peaks $V^-$ and $V^+$ only in the doped system, as shown enlarged in the inset. The red curve is shifted vertically for easier comparison. Left: Sketch of the ribbon suspended between tip and substrate with the tip retraction $z$ and the tip-\bb-unit distance $d_\text{\bb}$ indicated.
    \textbf{d} Lewis structure of a borylated ribbon segment and STM topography image of the \bb-GNR ($V=-\SI{300}{mV}$, $I=\SI{30}{pA}$). The red cross indicates the position from where the ribbon was lifted.
	}
\end{figure*}

We performed two-terminal electronic transport measurements in GNRs suspended between the tip and a Au(111) substrate of a low-temperature scanning tunneling microscope \citep{Koch2012, friedrich_magnetism_2020, friedrich_addressing_2022} as sketched in Fig.~\ref{fig:Fig1}\textbf{c}. To reach this configuration, the STM tip is positioned above one termination of the ribbon (e.g. over the red cross in Fig.~\ref{fig:Fig1}\textbf{d}), and gently approached towards the substrate until a bond between tip and ribbon forms. Upon retracting the STM tip, the ribbon is partially lifted from the substrate, creating the two-terminal transport configuration. Since the ribbons are semiconducting, their linear conductance for small voltage biases drops exponentially with increasing tip-substrate-distance $z$, with a  decay constant that is characteristic of co-tunneling through the GNR band-gap \citep{Koch2012}. 

To study the transport through a single \bb-dopant group, we retracted the ribbon to a tip height $z$ larger than \z2b, the spacing between the \bb\ site and GNR's end. Close to $z\sim \text{\z2b}$, the detachment of the 2B moiety from the surface is detected in STS spectra as a faint Kondo resonance \citep{friedrich_magnetism_2020}. A $\didv$ spectrum recorded above this point reveals two sharp peaks, $V^-$ and $V^+$ in Fig. \ref{fig:Fig1}\textbf{d}. They correspond to the onset of resonant band transport through the 7AGNR segments of the suspended ribbon as they are equally observed for a pristine 7AGNR lifted to a similar tip height (blue plot in Fig.~\ref{fig:Fig1}\textbf{d}).

In the \bb-GNRs, we observed additional spectral features between the band transport resonances. For the ribbon in Fig.~\ref{fig:Fig1}\textbf{d}, these in-gap features consist of two weaker $\didv$ peaks at $V_\text{2B}^-=\SI{-0.5}{V}$ and $V_\text{2B}^+=\SI{1.07}{V}$, shown in the figure's inset. At the retraction height of $z=\SI{5.8}{nm}>\text{\z2b}$, the \bb\ moiety lies in the free-standing segment of the ribbon and provides a exponentially localised state, which is decoupled from tip and substrate by the two pristine GNR segments and located in the ribbon's band gap (as in Fig.~\ref{fig:Fig1}\textbf{b}). In this configuration, electrical transport proceeds in a double-barrier fashion, where a fraction $\alpha V$ ($0 < \alpha < 1)$ of the tip-substrate bias $V$ drops between tip and \bb-state, and the rest (1-$\alpha)V$ between boron moiety and substrate, i.e., $\alpha$ quantifies the electrostatic coupling. The position of the $V_\text{2B}^{\pm}$ peaks in the spectra depends on the energy alignment of the \bb-states with respect to the chemical potential of the system ($\varepsilon$ and $\varepsilon'$ for the occupied and unoccupied state in Fig.~\ref{fig:Fig1}\textbf{b}) and on the value of $\alpha$. Specifically, $\alpha$, $\varepsilon$ and $\varepsilon'$ determine if the electrical transport is ambipolar (both O2B and U2B drive carriers at either polarity) or unipolar (resonant transport flow through only one of those states at either polarity) \citep{qiu_vibronic_2004, nazin_tunneling_2005}.

\begin{figure*}[t]
	\includegraphics[width=\textwidth]{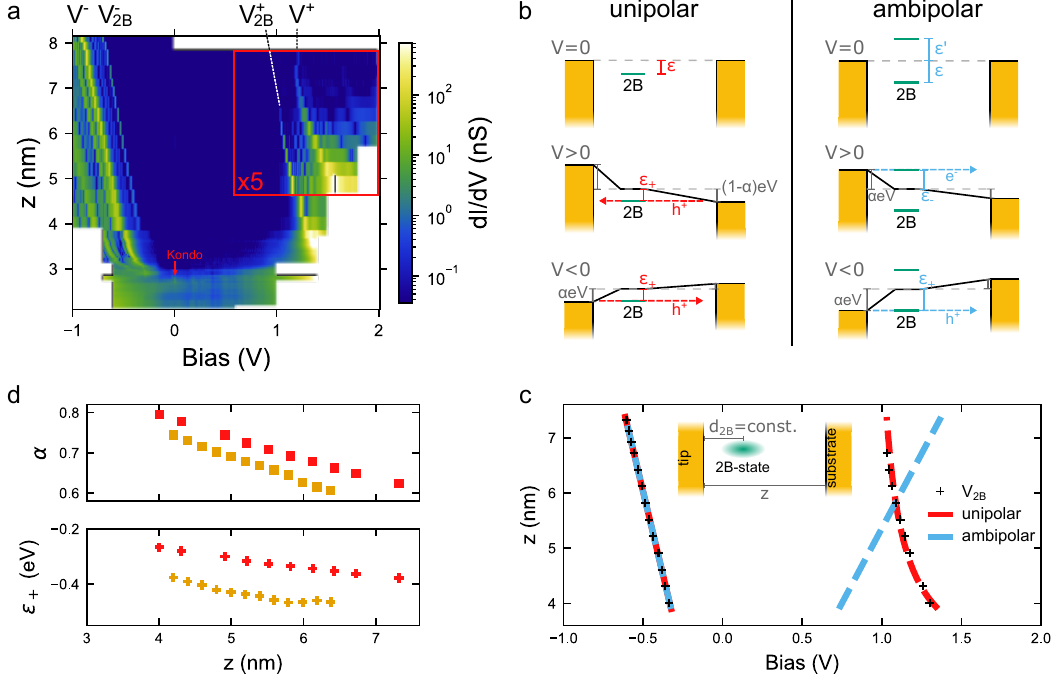}
	\caption{\label{fig:Fig2}
	\textbf{Resonant bipolar transport through the \bb-state:}
	\textbf{a} Colour map of $\didv$ spectra at different tip substrate separation $z$. The transport resonances at $V_\text{2B}^\pm$ and $V^\pm$ are clearly identified in the spectra for $z>\SI{3}{nm}$. The $\didv$-signal in the top right corner has been amplified by a factor $5$. A Kondo resonance (red arrow) emerges at $z=\SI{2.8}{nm}$ upon detaching the \bb-unit from the substrate \citep{friedrich_magnetism_2020}.
	\textbf{b} Comparison of unipolar transport through a single occupied energy level $\varepsilon$ and ambipolar transport through two energy levels $\varepsilon$ and $\varepsilon'$. The conditions for resonant transport through the O2B$^+$ hole resonance $\varepsilon_+$ (U2B$^-$ electron resonance $\varepsilon_-$) are indicated for positive and negative voltage in both cases. See SI for a more detailed discussion.
	\textbf{c} Simulated $V_\text{2B}^-$ and $V_\text{2B}^+$ as functions of $z$ for the unipolar model (red, dashed line) and the ambipolar model (blue, dashed line) assuming the plate capacitor geometry and the experimental values of $V_\text{2B}^-$ and $V_\text{2B}^+$ (black crosses) extracted from \textbf{a}. The inset sketches the plate capacitor model, from which follows $\alpha = d_\text{\bb}/z$. See SI for details on the simulation.
    \textbf{d} Gating efficiency $\alpha$ and energy $\varepsilon_+$ of the O2B$^+$ hole resonance for two \bb-GNR. The orange data points stem from a ribbon with the \bb-unit being $\sim \SI{0.9}{nm}$ closer to the tip, \textit{i.e.}, smaller $d_\text{\bb}$.
	}
\end{figure*}

To decipher the transport mechanisms behind the two in-gap resonances, we modified $\alpha$ by increasing the tip height, i.e. increasing the size of the \bb-substrate barrier, and studied changes in the $\didv$ spectra. In Fig.~\ref{fig:Fig2}\textbf{a}, we plot a set of $\didv$ spectra recorded in the range $\SI{2.1}{nm} < z < \SI{8.3}{nm}$ as a color map that pictures the peak evolution with $z$. The position of both peaks $V_\text{2B}^-$ and $V_\text{2B}^+$ shifts with $z$ towards more negative values, while their spacing is rather constant (see Supplementary Information (SI) for a detailed discussion). Note that the amplitude of $V_\text{2B}^-$ is fairly constant during retraction, while $V_\text{2B}^+$ decreases.

We modeled the $z$-dependency of the position of both $V_\text{2B}$ peaks by assuming that the fraction of potential drop $\alpha$ scales with the ratio $z$/\z2b, as in a parallel plate capacitor. Figure~\ref{fig:Fig2}\textbf{b} details the conditions for resonant transport in the different transport configurations, which lead to analytical expressions for the $z$-dependence of the $V_\text{2B}^-$ and $V_\text{2B}^+$ peaks (see SI for details). Using these expressions, we fitted in Fig.~\ref{fig:Fig2}\textbf{c} the peak positions $V_\text{2B}^-$ and $V_\text{2B}^+$ from Fig.~\ref{fig:Fig2}\textbf{a} with the ambipolar (blue line) and unipolar (red line) resonant transport model. Both models perfectly reproduce the $z$-dependence of $V_\text{2B}^-$, which represents hole transport \textit{via} the occupied \bb-state, i.e. the O2B$^+$ resonance. However, the energy shift of the $V_\text{2B}^+$ peak with $z$ is only reproduced in the case of unipolar transport, while the ambipolar behavior results in an opposite height dependence. The unipolar transport model even reproduces accurately a minute non-linearity observed only in the shift of $V_\text{2B}^+$. Therefore, we attribute both $V_\text{2B}^-$ and $V_\text{2B}^+$ peaks to unipolar hole transport through the O2B$^+$ resonance, and discard resonant electron tunneling through the (negative ion) state U2B$^-$.

The unipolar transport model also provides analytical expressions for the electrostatic coupling $\alpha = V_\text{2B}^+/(V_\text{2B}^+-V_\text{2B}^-)$ and the O2B$^+$ hole resonance $\varepsilon_+ = eV_\text{2B}^+V_\text{2B}^-/(V_\text{2B}^+ - V_\text{2B}^-)$, with the elementary charge $e$. We used these expressions to obtain the value of these two quantities at each value of $z$ from the experimental data available, without further assumptions (Fig.~\ref{fig:Fig2}\textbf{d}). We find that $\alpha$ decreases monotonously with increasing $z$. This agrees with the continuous increase of the \bb-surface barrier width, which increases $1-\alpha$, as the tip retracts \citep{Wagner2015sqdm}. Accordingly, a second GNR with the \bb-unit closer to the tip by $\sim \SI{0.9}{nm}$ (orange symbols in Fig.~\ref{fig:Fig2}\textbf{d}, refer to Fig.~S2 for the conductance map) shows a similar decay of $\alpha$ with $z$ but with a $\sim 5\%$ smaller value. Furthermore, the energy $\varepsilon_+$ of the O2B$^+$ hole resonance also shifts away from the Fermi level with increasing $z$ (Fig.~\ref{fig:Fig2}\textbf{d}), from $\sim \qtyrange[range-units = single]{-300}{-400}{meV}$. We attribute the shift of $\varepsilon_+$ to the reduction of the Au(111) surface-induced hole doping of the GNR \citep{Merino-Diez2018} as the ribbon is lifted from the surface \citep{lawrence_probing_2020}.

\begin{figure*}[t]
	\includegraphics[width=\textwidth]{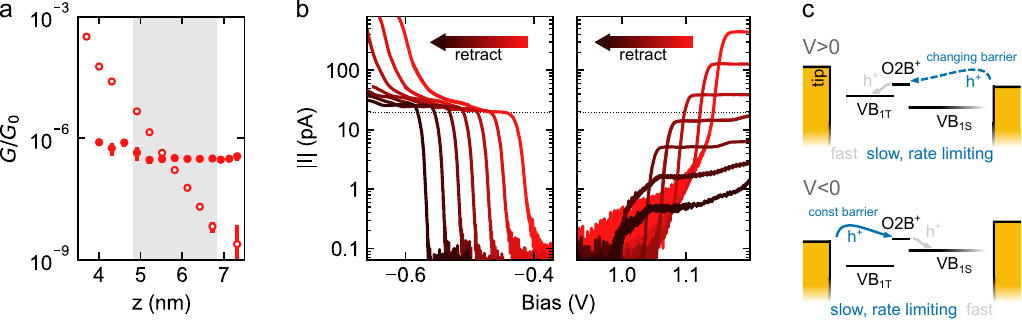}
	\caption{\label{fig:Fig3}
    \textbf{Current rectification through molecular asymmetry:}
    \textbf{a} Linear resonant conductance $G (V_\text{2B}^\pm, z)$ through the O2B$^+$ hole resonance at negative (filled circles) and positive bias (open circles). The decay constants $\beta^-_{>5} = -0.04\pm \SI{0.03}{nm^{-1}}$ (for $z>\SI{5}{nm}$) and $\beta^+ = 3.4 \pm \SI{0.1}{nm^{-1}}$ are obtained. The grey shaded region corresponds to the data presented in panel \textbf{b}.
    \textbf{b} $|I|$-$V$-curves taken from $z=\SI{4.9}{nm}$ (bright red curve) to $\SI{6.7}{nm}$ (dark red curve) in steps of $\Delta z =\SI{0.3}{nm}$. At negative bias, the onset current remains approximately constant. In stark contrast, the onset current at positive bias reduces exponentially. Note that the onset current at negative bias ($I\approx -\SI{20}{pA}$) slightly increases resulting in constant linear conductance.
	\textbf{c} Sketch of the VB-extension of the one-level unipolar transport model. Hole tunneling from tip ($V<0$) or substrate ($V>0$) to the O2B$^+$ hole resonance limits the tunneling rate, resulting in a constant ($z$-dependent) transmission for $V<0$ ($V>0$). The finite voltage drop across the \ce{2B}-state brings the valence band in the second ("grey") segment of the ribbon closer to the O2B$^+$ resonance, facilitating the fast hole relaxation into to the second electrode.
	}
\end{figure*}

\section{Current rectification}

A striking feature of the conductance map in Fig.~\ref{fig:Fig2}\textbf{a} is that, at negative voltage, the $\didv$ amplitude of O2B$^+$ hole resonance remains rather constant during the whole retraction, even for separations as large as $z=\SI{8}{nm}$. However, the amplitude of the resonance at positive bias voltage decreases monotonously with $z$. To quantify this, we compare in Fig.~\ref{fig:Fig3}\textbf{a} the linear conductance $G(V,z)$ of the $V_\text{2B}^\pm$ resonances. The values of the negative bias conductance $G(V_\text{2B}^-,z)$ (filled circles) remain fairly constant during the whole tip retraction. On the other hand, $G(V_\text{2B}^+,z)$ decreases exponentially (open circles), and changes by six orders of magnitude during the 4 nm retraction, with decay constant $\beta^+ = 3.4\pm\SI{0.1}{nm^{-1}}$. The ratio of resonant conductance $G(V_\text{2B}^{+}) / G(V_\text{2B}^{-})$ (defined as resonant RR) changes by six orders of magnitude over the observed $z$ interval (from $\sim 10^3$ to $10^{-3}$), showcasing a wide-range \textit{in-situ}-tunability of the GNR's diode character, exceeding other tunable devices \citep{batra_tuning_2013, perrin_gate-tunable_2016, atesci_humidity-controlled_2018}. Notably, the forward bias direction reverses at $z=\SI{5.7}{nm}$. At this elevation the diboron unit lies around the midpoint of the tip-surface gap. The inversion of rectification direction has so far only been achieved by chemical modifications to the employed molecule \citep{nguyen_control_2017}, further evidencing our extraordinary ability of mechanical control over the rectification behaviour of the molecular device.

The corresponding current-bias-traces recorded in the range $\SI{4.9}{nm} \leq z \leq \SI{6.7}{nm}$ (Fig.~\ref{fig:Fig3}\textbf{b}) reveal sharp current onsets upon reaching resonance conditions, followed by a constant current plateau. While for negative bias the current plateau amounts to $I\approx\SI{20}{pA}$ independently of $z$, at positive bias the current value is strongly sensitive to $z$. A non-resonant RR, defined as RR$_\text{nr}=|G(V)/G(-V)|$, quantifies the diode character of the borylated GNR in an alternative way. A RR$_\text{nr}$ in the range of $10^5 - 10^3$ is reached at low bias voltage ($\pm\SI{0.5}{V}$) and $z<\SI{5.5}{nm}$.

To rationalize the large current rectification in the unipolar transport \textit{via} the \bb-state, we consider the tunneling channel opened by the O2B$^+$ resonance (see Fig.~\ref{fig:Fig3}\textbf{c}). As evidenced in Fig.~\ref{fig:Fig2}, hole transport proceeds through the \bb-state when the positive-ion state O2B$^+$ aligns with either tip's ($V<0$) or substrate's ($V>0$) chemical potential. At negative bias, hole tunneling from the tip is the rate-limiting process (blue arrow in Fig.~\ref{fig:Fig3}\textbf{c} bottom), which remains constant with $z$ because the width \z2b\ of the tip-\bb\ barrier is fixed. At positive bias, the rate-limiting step is hole tunneling from the substrate through the GNR barrier (blue arrow in Fig.~\ref{fig:Fig3}\textbf{c} top), which increases with $z$. Therefore, conductance at positive bias decreases exponentially as the ribbon is lifted. The presence of the two segments of GNR valence band at $\sim \SI{200}{mV}$ below the \bb-state reduces the apparent height of the barrier for hole tunneling and enables tunneling over large distances \citep{Li2019, lawrence_probing_2020}. Additionally, a finite voltage drop across the \bb-state can alter the relative alignment of the two valence bands with respect to the O2B state, potentially lowering the height of the "second" tunneling barrier further and, thus, facilitating the fast relaxation of the hole into the second electrode at either bias polarity (grey arrows in Fig.~\ref{fig:Fig3}\textbf{c}).

\begin{figure*}[t]
	\includegraphics[width=\textwidth]{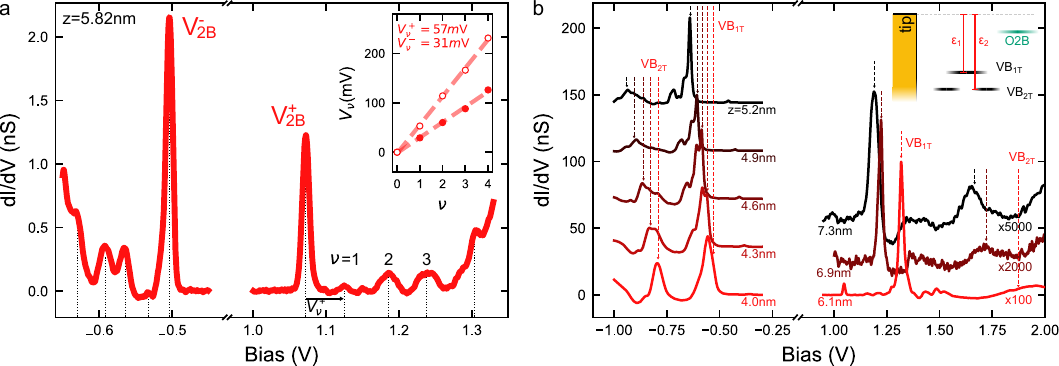}
	\caption{\label{fig:Fig4}
    \textbf{Phonon and Quantum Well Excitations:}
 	\textbf{a} $V_\text{2B}^-$ and $V_\text{2B}^+$ with their satellite structure. The equidistant spacing is the fingerprint of vibronic modes \citep{nazin_tunneling_2005, franke_effects_2012}. The inset shows the linear regression fitted to the relative peak voltages for negative (filled circles) and positive (open circles) bias. The obtained slopes are indicated. As the vibronic states undergo the same gating as O2B, the unipolar transport model can be applied. We obtain $\alpha = 0.65\pm 0.01$ and $E_\nu = 20.1\pm\SI{0.6}{meV}$. Note, that the relative peak intensities for both bias polarities are almost identical in agreement with unipolar transport.
    \textbf{b} $\didv$ spectra at selected $z$. The calculated voltages VB$_{2\text{T}}$ for quantum well resonances of the confined VB are indicated by vertical lines. Spectra are offset vertically for clarity. The inset sketches the first two quantum well states of the quantized VB in the particle-in-a-box model. 
	}
\end{figure*}

Elevating the bias beyond resonant transport through the \bb-state enables resonant band transport at $V^+$ and $V^-$. The $\didv$ peaks $V^+$ and $V^-$, and the peaks $V_\text{\bb}^+$ and $V_\text{\bb}^-$ exhibit essentially identical energy shift with $z$, evidencing that the band resonances found at higher bias correspond to unipolar resonant transport \textit{via} the VB ($V^\pm$ in Fig.~\ref{fig:Fig2}\textbf{a}, see also SI for a detailed discussion). Most importantly, this enables efficient current rectification for bias voltage up to $\SI{1}{V}$. The corresponding current plateau of $\SI{2}{nA}$ at negative bias voltage is accessible over a large $z$-interval and exhibits $\text{RR} > 10^6$, exceeding values of reported molecular diodes \citep{chen_molecular_2017, zhang_large_2021}.

\section{Resonant phonon and quantum well states}

The peaks $V_\text{2B}^-$ and $V_\text{2B}^+$ are accompanied by a set of satellites (Fig.~\ref{fig:Fig4}\textbf{a}) with equidistant energy spacing, which originate from unipolar resonant tunneling through vibronic states \citep{nazin_tunneling_2005, franke_effects_2012}. The features show a characteristic amplitude pattern indicating that the same Franck-Condon coefficients govern the vibronic modes independent of the voltage polarity. Fitting the satellite peak's energy by a linear regression (inset of Fig.~\ref{fig:Fig4}\textbf{a}), we obtain the excitation energy $eV_\nu^\pm$ at positive and negative bias.

The vibronic structure follows the same gating as the \bb-resonances. Therefore, we can apply the set of equations from the unipolar model discussed above to determine the fundamental vibrational energy $E_\nu$ and the corresponding $\alpha$ for every $z$. For example, we find that $E_\nu$ amounts to $20.1\pm\SI{0.6}{meV}$ and $\alpha = 0.65\pm 0.01$ at $z=\SI{5.82}{nm}$. We find that $\alpha$ decays with tip retraction in a similar fashion as for the \bb-states, while $E_\nu$ remains constant with $z$ (see SI, Fig.~S5). This excludes that $E_\nu$ corresponds to bending or torsional vibrations of the whole free standing segment. Instead, as we discuss in the SI (Fig.~S5 and S6), these peaks most probably correspond to a vibrational mode of the boron moiety. 

The confinement of the VB by the \bb-state gives rise to higher order quantum well modes inside the ribbon \citep{Carbonell-Sanroma2017} (Fig.~\ref{fig:Fig1}\textbf{b} and inset Fig.~\ref{fig:Fig4}\textbf{b}). Quantum well modes of 7AGNRs were spatially resolved by scanning tunneling spectroscopy \citep{Sode2015, Carbonell-Sanroma2017}. The sharp transport resonances $V^-$ and $V^+$, associated to the zero-order quantum well mode of the VB, are traced through a series of $\didv$ spectra of the \bgnr (VB$_{1\text{T}}$ in Fig.~\ref{fig:Fig4}\textbf{b}). Using a simple particle-in-a-box model (see SI for details), we estimate the voltage at which the transport resonance corresponding to the second quantum well mode VB$_{2\text{T}}$ between tip and \bb-state is expected (Fig.~\ref{fig:Fig4}\textbf{b}). Spectral resonances found above $V^-$ and $V^+$ in the $\didv$ spectra reproduce accurately the voltages obtained from the particle-in-a-box model confirming the access to quantum well modes of the confined VB \textit{via} unipolar resonant transport.

\section{Conclusion}
We have demonstrated the unipolar resonant transport through a boron-induced in-gap state embedded in a 7AGNR suspended between two metallic electrodes. The state is available for hole transport at both voltage polarities. This transport channel was predicted to show a large degree of spin polarization \citep{Zhang2022}. The asymmetric position of the quantum dot in the ribbon enables a large \textit{in situ} tunable current rectification (RR$>10^5$ at $V<\SI{1}{V}$) likely following a mechanism comparable to current rectification in asymmetric two-level molecules \citep{Elbing2005, perrin_large_2014, perrin_gate-tunable_2016}. We are able to mechanically control the resonant RR by six orders of magnitude, including an inversion of forward bias direction. Furthermore, the presence of the boron moiety confines the valence band, creating quantum well states in the ribbon. We show that both ground and excited states of the quantized band are available for resonant electron transport. Our results extend the technological potential of atomically precise high-conductance graphene nanoribbons and shine light on a novel approach for the manipulation of molecular electronic functionalities.

\section{Methods}
The Au(111) substrate was prepared by sputtering the crystal with Neon ions for 10 minutes, followed by one annealing step at $T=\SI{740}{K}$ in ultra-high vacuum conditions ($p<\SI{e-9}{mbar}$) for 15 minutes.
The ribbons were fabricated following the two-step on-surface synthesis \textit{via} successive Ullmann-coupling and cyclodehydrogenation \citep{Carbonell-Sanroma2017, Carbonell-Sanroma2018, friedrich_magnetism_2020}. The samples are analyzed \textit{in situ} in an homebuild STM kept with liquid Helium at $\sim \SI{6}{K}$. We use a PtIr tip. An atomically sharp apex termination by gold atoms was reached by controlled indention of the tip into the clean Au(111) substrate.
Details on the lifting procedure to reach the two-terminal transport configuration are described in detail elsewhere \citep{friedrich_magnetism_2020}.
The experimental data was prepared for presentation using WSxM \citep{horcas_wsxm_2007} and the python matplotlib libary \citep{hunter_matplotlib_2007}. Colour maps use preceptual continuous colour scales based on \citep{kovesi_good_2015}.

%

\section*{Acknowledgements}
We gratefully acknowledge financial support from Spanish MCIN/AEI/ 10.13039/501100011033 and the European Regional Development Fund (ERDF) through grants with number PID2019-107338RBC6, PID2022-140845OB-C6, and CEX2020-001038-M,  and from the European Union (EU) through Horizon 2020 (FET-Open project SPRING Grant.~no.~863098).

\section*{Author information}
\subsection*{Contributions}
NF and JIP conceived the experiment. 
NF and JL performed the on-sufrace synthesis and STM/STS measurments. 
IP and DP synthesized the molecular precursors. 
NF analysed the experimental data and developed the theoretical model.
NF, JL and JIP discussed the data and model.
NF and JIP wrote the manuscript. 
All authors discussed and approved the manuscript.

\section{Ethics Declaration}
The authors declare no competing interests.


\begin{thebibliography}{47}%
\makeatletter
\providecommand \@ifxundefined [1]{%
 \@ifx{#1\undefined}
}%
\providecommand \@ifnum [1]{%
 \ifnum #1\expandafter \@firstoftwo
 \else \expandafter \@secondoftwo
 \fi
}%
\providecommand \@ifx [1]{%
 \ifx #1\expandafter \@firstoftwo
 \else \expandafter \@secondoftwo
 \fi
}%
\providecommand \natexlab [1]{#1}%
\providecommand \enquote  [1]{``#1''}%
\providecommand \bibnamefont  [1]{#1}%
\providecommand \bibfnamefont [1]{#1}%
\providecommand \citenamefont [1]{#1}%
\providecommand \href@noop [0]{\@secondoftwo}%
\providecommand \href [0]{\begingroup \@sanitize@url \@href}%
\providecommand \@href[1]{\@@startlink{#1}\@@href}%
\providecommand \@@href[1]{\endgroup#1\@@endlink}%
\providecommand \@sanitize@url [0]{\catcode `\\12\catcode `\$12\catcode
  `\&12\catcode `\#12\catcode `\^12\catcode `\_12\catcode `\%12\relax}%
\providecommand \@@startlink[1]{}%
\providecommand \@@endlink[0]{}%
\providecommand \url  [0]{\begingroup\@sanitize@url \@url }%
\providecommand \@url [1]{\endgroup\@href {#1}{\urlprefix }}%
\providecommand \urlprefix  [0]{URL }%
\providecommand \Eprint [0]{\href }%
\providecommand \doibase [0]{http://dx.doi.org/}%
\providecommand \selectlanguage [0]{\@gobble}%
\providecommand \bibinfo  [0]{\@secondoftwo}%
\providecommand \bibfield  [0]{\@secondoftwo}%
\providecommand \translation [1]{[#1]}%
\providecommand \BibitemOpen [0]{}%
\providecommand \bibitemStop [0]{}%
\providecommand \bibitemNoStop [0]{.\EOS\space}%
\providecommand \EOS [0]{\spacefactor3000\relax}%
\providecommand \BibitemShut  [1]{\csname bibitem#1\endcsname}%
\let\auto@bib@innerbib\@empty
\bibitem [{\citenamefont {Aviram}\ and\ \citenamefont
  {Ratner}(1974)}]{Aviram1974}%
  \BibitemOpen
  \bibfield  {author} {\bibinfo {author} {\bibfnamefont {A.}~\bibnamefont
  {Aviram}}\ and\ \bibinfo {author} {\bibfnamefont {M.~A.}\ \bibnamefont
  {Ratner}},\ }\href {\doibase 10.1016/0009-2614(74)85031-1} {\bibfield
  {journal} {\bibinfo  {journal} {Chemical physics letters}\ }\textbf {\bibinfo
  {volume} {29}},\ \bibinfo {pages} {277} (\bibinfo {year} {1974})}\BibitemShut
  {NoStop}%
\bibitem [{\citenamefont {Gupta}\ \emph {et~al.}(2023)\citenamefont {Gupta},
  \citenamefont {Fereiro}, \citenamefont {Bayat}, \citenamefont {Pritam},
  \citenamefont {Zharnikov},\ and\ \citenamefont
  {Mondal}}]{gupta_nanoscale_2023}%
  \BibitemOpen
  \bibfield  {author} {\bibinfo {author} {\bibfnamefont {R.}~\bibnamefont
  {Gupta}}, \bibinfo {author} {\bibfnamefont {J.~A.}\ \bibnamefont {Fereiro}},
  \bibinfo {author} {\bibfnamefont {A.}~\bibnamefont {Bayat}}, \bibinfo
  {author} {\bibfnamefont {A.}~\bibnamefont {Pritam}}, \bibinfo {author}
  {\bibfnamefont {M.}~\bibnamefont {Zharnikov}}, \ and\ \bibinfo {author}
  {\bibfnamefont {P.~C.}\ \bibnamefont {Mondal}},\ }\href {\doibase
  10.1038/s41570-022-00457-8} {\bibfield  {journal} {\bibinfo  {journal} {Nat.
  Rev. Chem.}\ }\textbf {\bibinfo {volume} {7}},\ \bibinfo {pages} {106}
  (\bibinfo {year} {2023})}\BibitemShut {NoStop}%
\bibitem [{\citenamefont {Yamada}\ \emph {et~al.}(2008)\citenamefont {Yamada},
  \citenamefont {Kumazawa}, \citenamefont {Noutoshi}, \citenamefont {Tanaka},\
  and\ \citenamefont {Tada}}]{yamada_electrical_2008}%
  \BibitemOpen
  \bibfield  {author} {\bibinfo {author} {\bibfnamefont {R.}~\bibnamefont
  {Yamada}}, \bibinfo {author} {\bibfnamefont {H.}~\bibnamefont {Kumazawa}},
  \bibinfo {author} {\bibfnamefont {T.}~\bibnamefont {Noutoshi}}, \bibinfo
  {author} {\bibfnamefont {S.}~\bibnamefont {Tanaka}}, \ and\ \bibinfo {author}
  {\bibfnamefont {H.}~\bibnamefont {Tada}},\ }\href {\doibase
  10.1021/nl0732023} {\bibfield  {journal} {\bibinfo  {journal} {Nano Lett.}\
  }\textbf {\bibinfo {volume} {8}},\ \bibinfo {pages} {1237} (\bibinfo {year}
  {2008})}\BibitemShut {NoStop}%
\bibitem [{\citenamefont {Gehring}\ \emph {et~al.}(2019)\citenamefont
  {Gehring}, \citenamefont {Thijssen},\ and\ \citenamefont {van~der
  Zant}}]{gehring_single-molecule_2019}%
  \BibitemOpen
  \bibfield  {author} {\bibinfo {author} {\bibfnamefont {P.}~\bibnamefont
  {Gehring}}, \bibinfo {author} {\bibfnamefont {J.~M.}\ \bibnamefont
  {Thijssen}}, \ and\ \bibinfo {author} {\bibfnamefont {H.~S.~J.}\ \bibnamefont
  {van~der Zant}},\ }\href {\doibase 10.1038/s42254-019-0055-1} {\bibfield
  {journal} {\bibinfo  {journal} {Nat. Rev. Phys.}\ }\textbf {\bibinfo {volume}
  {1}},\ \bibinfo {pages} {381} (\bibinfo {year} {2019})}\BibitemShut {NoStop}%
\bibitem [{\citenamefont {Chen}\ \emph
  {et~al.}(2017{\natexlab{a}})\citenamefont {Chen}, \citenamefont {Roemer},
  \citenamefont {Yuan}, \citenamefont {Du}, \citenamefont {Thompson},
  \citenamefont {Del~Barco},\ and\ \citenamefont {Nijhuis}}]{Chen2017}%
  \BibitemOpen
  \bibfield  {author} {\bibinfo {author} {\bibfnamefont {X.}~\bibnamefont
  {Chen}}, \bibinfo {author} {\bibfnamefont {M.}~\bibnamefont {Roemer}},
  \bibinfo {author} {\bibfnamefont {L.}~\bibnamefont {Yuan}}, \bibinfo {author}
  {\bibfnamefont {W.}~\bibnamefont {Du}}, \bibinfo {author} {\bibfnamefont
  {D.}~\bibnamefont {Thompson}}, \bibinfo {author} {\bibfnamefont
  {E.}~\bibnamefont {Del~Barco}}, \ and\ \bibinfo {author} {\bibfnamefont
  {C.~A.}\ \bibnamefont {Nijhuis}},\ }\href {\doibase 10.1038/nnano.2017.110}
  {\bibfield  {journal} {\bibinfo  {journal} {Nature Nanotechnol.}\ }\textbf
  {\bibinfo {volume} {12}},\ \bibinfo {pages} {797} (\bibinfo {year}
  {2017}{\natexlab{a}})}\BibitemShut {NoStop}%
\bibitem [{\citenamefont {Atesci}\ \emph {et~al.}(2018)\citenamefont {Atesci},
  \citenamefont {Kaliginedi}, \citenamefont {Celis~Gil}, \citenamefont {Ozawa},
  \citenamefont {Thijssen}, \citenamefont {Broekmann}, \citenamefont {Haga},\
  and\ \citenamefont {van~der Molen}}]{atesci_humidity-controlled_2018}%
  \BibitemOpen
  \bibfield  {author} {\bibinfo {author} {\bibfnamefont {H.}~\bibnamefont
  {Atesci}}, \bibinfo {author} {\bibfnamefont {V.}~\bibnamefont {Kaliginedi}},
  \bibinfo {author} {\bibfnamefont {J.~A.}\ \bibnamefont {Celis~Gil}}, \bibinfo
  {author} {\bibfnamefont {H.}~\bibnamefont {Ozawa}}, \bibinfo {author}
  {\bibfnamefont {J.~M.}\ \bibnamefont {Thijssen}}, \bibinfo {author}
  {\bibfnamefont {P.}~\bibnamefont {Broekmann}}, \bibinfo {author}
  {\bibfnamefont {M.-a.}\ \bibnamefont {Haga}}, \ and\ \bibinfo {author}
  {\bibfnamefont {S.~J.}\ \bibnamefont {van~der Molen}},\ }\href {\doibase
  10.1038/s41565-017-0016-8} {\bibfield  {journal} {\bibinfo  {journal} {Nat.
  Nanotechnol.}\ }\textbf {\bibinfo {volume} {13}},\ \bibinfo {pages} {117}
  (\bibinfo {year} {2018})}\BibitemShut {NoStop}%
\bibitem [{\citenamefont {Batra}\ \emph {et~al.}(2013)\citenamefont {Batra},
  \citenamefont {Darancet}, \citenamefont {Chen}, \citenamefont {Meisner},
  \citenamefont {Widawsky}, \citenamefont {Neaton}, \citenamefont {Nuckolls},\
  and\ \citenamefont {Venkataraman}}]{batra_tuning_2013}%
  \BibitemOpen
  \bibfield  {author} {\bibinfo {author} {\bibfnamefont {A.}~\bibnamefont
  {Batra}}, \bibinfo {author} {\bibfnamefont {P.}~\bibnamefont {Darancet}},
  \bibinfo {author} {\bibfnamefont {Q.}~\bibnamefont {Chen}}, \bibinfo {author}
  {\bibfnamefont {J.~S.}\ \bibnamefont {Meisner}}, \bibinfo {author}
  {\bibfnamefont {J.~R.}\ \bibnamefont {Widawsky}}, \bibinfo {author}
  {\bibfnamefont {J.~B.}\ \bibnamefont {Neaton}}, \bibinfo {author}
  {\bibfnamefont {C.}~\bibnamefont {Nuckolls}}, \ and\ \bibinfo {author}
  {\bibfnamefont {L.}~\bibnamefont {Venkataraman}},\ }\href {\doibase
  10.1021/nl403698m} {\bibfield  {journal} {\bibinfo  {journal} {Nano Lett.}\
  }\textbf {\bibinfo {volume} {13}},\ \bibinfo {pages} {6233} (\bibinfo {year}
  {2013})}\BibitemShut {NoStop}%
\bibitem [{\citenamefont {Xin}\ \emph {et~al.}(2017)\citenamefont {Xin},
  \citenamefont {Wang}, \citenamefont {Jia}, \citenamefont {Liu}, \citenamefont
  {Zhang}, \citenamefont {Yu}, \citenamefont {Li}, \citenamefont {Wang},
  \citenamefont {Gong}, \citenamefont {Sun}, \citenamefont {Zhang},
  \citenamefont {Liu}, \citenamefont {Zhang}, \citenamefont {Liao},
  \citenamefont {Zhang},\ and\ \citenamefont
  {Guo}}]{xin_stereoelectronic_2017}%
  \BibitemOpen
  \bibfield  {author} {\bibinfo {author} {\bibfnamefont {N.}~\bibnamefont
  {Xin}}, \bibinfo {author} {\bibfnamefont {J.}~\bibnamefont {Wang}}, \bibinfo
  {author} {\bibfnamefont {C.}~\bibnamefont {Jia}}, \bibinfo {author}
  {\bibfnamefont {Z.}~\bibnamefont {Liu}}, \bibinfo {author} {\bibfnamefont
  {X.}~\bibnamefont {Zhang}}, \bibinfo {author} {\bibfnamefont
  {C.}~\bibnamefont {Yu}}, \bibinfo {author} {\bibfnamefont {M.}~\bibnamefont
  {Li}}, \bibinfo {author} {\bibfnamefont {S.}~\bibnamefont {Wang}}, \bibinfo
  {author} {\bibfnamefont {Y.}~\bibnamefont {Gong}}, \bibinfo {author}
  {\bibfnamefont {H.}~\bibnamefont {Sun}}, \bibinfo {author} {\bibfnamefont
  {G.}~\bibnamefont {Zhang}}, \bibinfo {author} {\bibfnamefont
  {Z.}~\bibnamefont {Liu}}, \bibinfo {author} {\bibfnamefont {G.}~\bibnamefont
  {Zhang}}, \bibinfo {author} {\bibfnamefont {J.}~\bibnamefont {Liao}},
  \bibinfo {author} {\bibfnamefont {D.}~\bibnamefont {Zhang}}, \ and\ \bibinfo
  {author} {\bibfnamefont {X.}~\bibnamefont {Guo}},\ }\href {\doibase
  10.1021/acs.nanolett.6b04139} {\bibfield  {journal} {\bibinfo  {journal}
  {Nano Letters}\ }\textbf {\bibinfo {volume} {17}},\ \bibinfo {pages} {856}
  (\bibinfo {year} {2017})}\BibitemShut {NoStop}%
\bibitem [{\citenamefont {Nguyen}\ \emph {et~al.}(2017)\citenamefont {Nguyen},
  \citenamefont {Martin}, \citenamefont {Frath}, \citenamefont {Della~Rocca},
  \citenamefont {Lafolet}, \citenamefont {Barraud}, \citenamefont {Lafarge},
  \citenamefont {Mukundan}, \citenamefont {James}, \citenamefont {McCreery},\
  and\ \citenamefont {Lacroix}}]{nguyen_control_2017}%
  \BibitemOpen
  \bibfield  {author} {\bibinfo {author} {\bibfnamefont {Q.~v.}\ \bibnamefont
  {Nguyen}}, \bibinfo {author} {\bibfnamefont {P.}~\bibnamefont {Martin}},
  \bibinfo {author} {\bibfnamefont {D.}~\bibnamefont {Frath}}, \bibinfo
  {author} {\bibfnamefont {M.~L.}\ \bibnamefont {Della~Rocca}}, \bibinfo
  {author} {\bibfnamefont {F.}~\bibnamefont {Lafolet}}, \bibinfo {author}
  {\bibfnamefont {C.}~\bibnamefont {Barraud}}, \bibinfo {author} {\bibfnamefont
  {P.}~\bibnamefont {Lafarge}}, \bibinfo {author} {\bibfnamefont
  {V.}~\bibnamefont {Mukundan}}, \bibinfo {author} {\bibfnamefont
  {D.}~\bibnamefont {James}}, \bibinfo {author} {\bibfnamefont {R.~L.}\
  \bibnamefont {McCreery}}, \ and\ \bibinfo {author} {\bibfnamefont {J.-C.}\
  \bibnamefont {Lacroix}},\ }\href {\doibase 10.1021/jacs.7b05732} {\bibfield
  {journal} {\bibinfo  {journal} {J. Am. Chem. Soc.}\ }\textbf {\bibinfo
  {volume} {139}},\ \bibinfo {pages} {11913} (\bibinfo {year}
  {2017})}\BibitemShut {NoStop}%
\bibitem [{\citenamefont {Yuan}\ \emph {et~al.}(2015)\citenamefont {Yuan},
  \citenamefont {Nerngchamnong}, \citenamefont {Cao}, \citenamefont {Hamoudi},
  \citenamefont {Del~Barco}, \citenamefont {Roemer}, \citenamefont {Sriramula},
  \citenamefont {Thompson},\ and\ \citenamefont
  {Nijhuis}}]{yuan_controlling_2015}%
  \BibitemOpen
  \bibfield  {author} {\bibinfo {author} {\bibfnamefont {L.}~\bibnamefont
  {Yuan}}, \bibinfo {author} {\bibfnamefont {N.}~\bibnamefont {Nerngchamnong}},
  \bibinfo {author} {\bibfnamefont {L.}~\bibnamefont {Cao}}, \bibinfo {author}
  {\bibfnamefont {H.}~\bibnamefont {Hamoudi}}, \bibinfo {author} {\bibfnamefont
  {E.}~\bibnamefont {Del~Barco}}, \bibinfo {author} {\bibfnamefont
  {M.}~\bibnamefont {Roemer}}, \bibinfo {author} {\bibfnamefont {R.~K.}\
  \bibnamefont {Sriramula}}, \bibinfo {author} {\bibfnamefont {D.}~\bibnamefont
  {Thompson}}, \ and\ \bibinfo {author} {\bibfnamefont {C.~A.}\ \bibnamefont
  {Nijhuis}},\ }\href {\doibase 10.1038/ncomms7324} {\bibfield  {journal}
  {\bibinfo  {journal} {Nat. Commun.}\ }\textbf {\bibinfo {volume} {6}},\
  \bibinfo {pages} {6324} (\bibinfo {year} {2015})}\BibitemShut {NoStop}%
\bibitem [{\citenamefont {Clair}\ and\ \citenamefont
  {de~Oteyza}(2019)}]{Clair2019}%
  \BibitemOpen
  \bibfield  {author} {\bibinfo {author} {\bibfnamefont {S.}~\bibnamefont
  {Clair}}\ and\ \bibinfo {author} {\bibfnamefont {D.~G.}\ \bibnamefont
  {de~Oteyza}},\ }\href {\doibase 10.1021/acs.chemrev.8b00601} {\bibfield
  {journal} {\bibinfo  {journal} {Chemical reviews}\ }\textbf {\bibinfo
  {volume} {119}},\ \bibinfo {pages} {4717} (\bibinfo {year}
  {2019})}\BibitemShut {NoStop}%
\bibitem [{\citenamefont {Cai}\ \emph {et~al.}(2014)\citenamefont {Cai},
  \citenamefont {Pignedoli}, \citenamefont {Talirz}, \citenamefont {Ruffieux},
  \citenamefont {Söde}, \citenamefont {Liang}, \citenamefont {Meunier},
  \citenamefont {Berger}, \citenamefont {Li}, \citenamefont {Feng},
  \citenamefont {Müllen},\ and\ \citenamefont {Fasel}}]{cai_graphene_2014}%
  \BibitemOpen
  \bibfield  {author} {\bibinfo {author} {\bibfnamefont {J.}~\bibnamefont
  {Cai}}, \bibinfo {author} {\bibfnamefont {C.~A.}\ \bibnamefont {Pignedoli}},
  \bibinfo {author} {\bibfnamefont {L.}~\bibnamefont {Talirz}}, \bibinfo
  {author} {\bibfnamefont {P.}~\bibnamefont {Ruffieux}}, \bibinfo {author}
  {\bibfnamefont {H.}~\bibnamefont {Söde}}, \bibinfo {author} {\bibfnamefont
  {L.}~\bibnamefont {Liang}}, \bibinfo {author} {\bibfnamefont
  {V.}~\bibnamefont {Meunier}}, \bibinfo {author} {\bibfnamefont
  {R.}~\bibnamefont {Berger}}, \bibinfo {author} {\bibfnamefont
  {R.}~\bibnamefont {Li}}, \bibinfo {author} {\bibfnamefont {X.}~\bibnamefont
  {Feng}}, \bibinfo {author} {\bibfnamefont {K.}~\bibnamefont {Müllen}}, \
  and\ \bibinfo {author} {\bibfnamefont {R.}~\bibnamefont {Fasel}},\ }\href
  {\doibase 10.1038/nnano.2014.184} {\bibfield  {journal} {\bibinfo  {journal}
  {Nat. Nanotechnol.}\ }\textbf {\bibinfo {volume} {9}},\ \bibinfo {pages}
  {896} (\bibinfo {year} {2014})}\BibitemShut {NoStop}%
\bibitem [{\citenamefont {Chen}\ \emph {et~al.}(2015)\citenamefont {Chen},
  \citenamefont {Cao}, \citenamefont {Chen}, \citenamefont {Pedramrazi},
  \citenamefont {Haberer}, \citenamefont {de~Oteyza}, \citenamefont {Fischer},
  \citenamefont {Louie},\ and\ \citenamefont {Crommie}}]{chen_molecular_2015}%
  \BibitemOpen
  \bibfield  {author} {\bibinfo {author} {\bibfnamefont {Y.-C.}\ \bibnamefont
  {Chen}}, \bibinfo {author} {\bibfnamefont {T.}~\bibnamefont {Cao}}, \bibinfo
  {author} {\bibfnamefont {C.}~\bibnamefont {Chen}}, \bibinfo {author}
  {\bibfnamefont {Z.}~\bibnamefont {Pedramrazi}}, \bibinfo {author}
  {\bibfnamefont {D.}~\bibnamefont {Haberer}}, \bibinfo {author} {\bibfnamefont
  {D.~G.}\ \bibnamefont {de~Oteyza}}, \bibinfo {author} {\bibfnamefont {F.~R.}\
  \bibnamefont {Fischer}}, \bibinfo {author} {\bibfnamefont {S.~G.}\
  \bibnamefont {Louie}}, \ and\ \bibinfo {author} {\bibfnamefont {M.~F.}\
  \bibnamefont {Crommie}},\ }\href {\doibase 10.1038/nnano.2014.307} {\bibfield
   {journal} {\bibinfo  {journal} {Nat. Nanotechnol.}\ }\textbf {\bibinfo
  {volume} {10}},\ \bibinfo {pages} {156} (\bibinfo {year} {2015})}\BibitemShut
  {NoStop}%
\bibitem [{\citenamefont {Carbonell-Sanromà}\ \emph
  {et~al.}(2017{\natexlab{a}})\citenamefont {Carbonell-Sanromà}, \citenamefont
  {Brandimarte}, \citenamefont {Balog}, \citenamefont {Corso}, \citenamefont
  {Kawai}, \citenamefont {Garcia-Lekue}, \citenamefont {Saito}, \citenamefont
  {Yamaguchi}, \citenamefont {Meyer}, \citenamefont {Sánchez-Portal},\ and\
  \citenamefont {Pascual}}]{Carbonell-Sanroma2017}%
  \BibitemOpen
  \bibfield  {author} {\bibinfo {author} {\bibfnamefont {E.}~\bibnamefont
  {Carbonell-Sanromà}}, \bibinfo {author} {\bibfnamefont {P.}~\bibnamefont
  {Brandimarte}}, \bibinfo {author} {\bibfnamefont {R.}~\bibnamefont {Balog}},
  \bibinfo {author} {\bibfnamefont {M.}~\bibnamefont {Corso}}, \bibinfo
  {author} {\bibfnamefont {S.}~\bibnamefont {Kawai}}, \bibinfo {author}
  {\bibfnamefont {A.}~\bibnamefont {Garcia-Lekue}}, \bibinfo {author}
  {\bibfnamefont {S.}~\bibnamefont {Saito}}, \bibinfo {author} {\bibfnamefont
  {S.}~\bibnamefont {Yamaguchi}}, \bibinfo {author} {\bibfnamefont
  {E.}~\bibnamefont {Meyer}}, \bibinfo {author} {\bibfnamefont
  {D.}~\bibnamefont {Sánchez-Portal}}, \ and\ \bibinfo {author} {\bibfnamefont
  {J.~I.}\ \bibnamefont {Pascual}},\ }\href {\doibase
  10.1021/acs.nanolett.6b03148} {\bibfield  {journal} {\bibinfo  {journal}
  {Nano Lett.}\ }\textbf {\bibinfo {volume} {17}},\ \bibinfo {pages} {50}
  (\bibinfo {year} {2017}{\natexlab{a}})}\BibitemShut {NoStop}%
\bibitem [{\citenamefont {Rizzo}\ \emph {et~al.}(2021)\citenamefont {Rizzo},
  \citenamefont {Jiang}, \citenamefont {Joshi}, \citenamefont {Veber},
  \citenamefont {Bronner}, \citenamefont {Durr}, \citenamefont {Jacobse},
  \citenamefont {Cao}, \citenamefont {Kalayjian}, \citenamefont {Rodriguez},
  \citenamefont {Butler}, \citenamefont {Chen}, \citenamefont {Louie},
  \citenamefont {Fischer},\ and\ \citenamefont
  {Crommie}}]{rizzo_rationally_2021}%
  \BibitemOpen
  \bibfield  {author} {\bibinfo {author} {\bibfnamefont {D.~J.}\ \bibnamefont
  {Rizzo}}, \bibinfo {author} {\bibfnamefont {J.}~\bibnamefont {Jiang}},
  \bibinfo {author} {\bibfnamefont {D.}~\bibnamefont {Joshi}}, \bibinfo
  {author} {\bibfnamefont {G.}~\bibnamefont {Veber}}, \bibinfo {author}
  {\bibfnamefont {C.}~\bibnamefont {Bronner}}, \bibinfo {author} {\bibfnamefont
  {R.~A.}\ \bibnamefont {Durr}}, \bibinfo {author} {\bibfnamefont {P.~H.}\
  \bibnamefont {Jacobse}}, \bibinfo {author} {\bibfnamefont {T.}~\bibnamefont
  {Cao}}, \bibinfo {author} {\bibfnamefont {A.}~\bibnamefont {Kalayjian}},
  \bibinfo {author} {\bibfnamefont {H.}~\bibnamefont {Rodriguez}}, \bibinfo
  {author} {\bibfnamefont {P.}~\bibnamefont {Butler}}, \bibinfo {author}
  {\bibfnamefont {T.}~\bibnamefont {Chen}}, \bibinfo {author} {\bibfnamefont
  {S.~G.}\ \bibnamefont {Louie}}, \bibinfo {author} {\bibfnamefont {F.~R.}\
  \bibnamefont {Fischer}}, \ and\ \bibinfo {author} {\bibfnamefont {M.~F.}\
  \bibnamefont {Crommie}},\ }\href {\doibase 10.1021/acsnano.1c09503}
  {\bibfield  {journal} {\bibinfo  {journal} {ACS Nano}\ }\textbf {\bibinfo
  {volume} {15}},\ \bibinfo {pages} {20633} (\bibinfo {year}
  {2021})}\BibitemShut {NoStop}%
\bibitem [{\citenamefont {Li}\ \emph {et~al.}(2018)\citenamefont {Li},
  \citenamefont {Merino-Díez}, \citenamefont {Carbonell-Sanromà},
  \citenamefont {Vilas-Varela}, \citenamefont {de~Oteyza}, \citenamefont
  {Peña}, \citenamefont {Corso},\ and\ \citenamefont {Pascual}}]{Li2018}%
  \BibitemOpen
  \bibfield  {author} {\bibinfo {author} {\bibfnamefont {J.}~\bibnamefont
  {Li}}, \bibinfo {author} {\bibfnamefont {N.}~\bibnamefont {Merino-Díez}},
  \bibinfo {author} {\bibfnamefont {E.}~\bibnamefont {Carbonell-Sanromà}},
  \bibinfo {author} {\bibfnamefont {M.}~\bibnamefont {Vilas-Varela}}, \bibinfo
  {author} {\bibfnamefont {D.~G.}\ \bibnamefont {de~Oteyza}}, \bibinfo {author}
  {\bibfnamefont {D.}~\bibnamefont {Peña}}, \bibinfo {author} {\bibfnamefont
  {M.}~\bibnamefont {Corso}}, \ and\ \bibinfo {author} {\bibfnamefont {J.~I.}\
  \bibnamefont {Pascual}},\ }\href {\doibase 10.1126/sciadv.aaq0582} {\bibfield
   {journal} {\bibinfo  {journal} {Sci. Adv.}\ }\textbf {\bibinfo {volume}
  {4}},\ \bibinfo {pages} {eaaq0582} (\bibinfo {year} {2018})}\BibitemShut
  {NoStop}%
\bibitem [{\citenamefont {Friedrich}\ \emph {et~al.}(2022)\citenamefont
  {Friedrich}, \citenamefont {Menchón}, \citenamefont {Pozo}, \citenamefont
  {Hieulle}, \citenamefont {Vegliante}, \citenamefont {Li}, \citenamefont
  {Sánchez-Portal}, \citenamefont {Peña}, \citenamefont {Garcia-Lekue},\ and\
  \citenamefont {Pascual}}]{friedrich_addressing_2022}%
  \BibitemOpen
  \bibfield  {author} {\bibinfo {author} {\bibfnamefont {N.}~\bibnamefont
  {Friedrich}}, \bibinfo {author} {\bibfnamefont {R.~E.}\ \bibnamefont
  {Menchón}}, \bibinfo {author} {\bibfnamefont {I.}~\bibnamefont {Pozo}},
  \bibinfo {author} {\bibfnamefont {J.}~\bibnamefont {Hieulle}}, \bibinfo
  {author} {\bibfnamefont {A.}~\bibnamefont {Vegliante}}, \bibinfo {author}
  {\bibfnamefont {J.}~\bibnamefont {Li}}, \bibinfo {author} {\bibfnamefont
  {D.}~\bibnamefont {Sánchez-Portal}}, \bibinfo {author} {\bibfnamefont
  {D.}~\bibnamefont {Peña}}, \bibinfo {author} {\bibfnamefont
  {A.}~\bibnamefont {Garcia-Lekue}}, \ and\ \bibinfo {author} {\bibfnamefont
  {J.~I.}\ \bibnamefont {Pascual}},\ }\href {\doibase 10.1021/acsnano.2c05673}
  {\bibfield  {journal} {\bibinfo  {journal} {ACS Nano}\ }\textbf {\bibinfo
  {volume} {16}},\ \bibinfo {pages} {14819} (\bibinfo {year}
  {2022})}\BibitemShut {NoStop}%
\bibitem [{\citenamefont {Senkovskiy}\ \emph {et~al.}(2017)\citenamefont
  {Senkovskiy}, \citenamefont {Pfeiffer}, \citenamefont {Alavi}, \citenamefont
  {Bliesener}, \citenamefont {Zhu}, \citenamefont {Michel}, \citenamefont
  {Fedorov}, \citenamefont {German}, \citenamefont {Hertel}, \citenamefont
  {Haberer}, \citenamefont {Petaccia}, \citenamefont {Fischer}, \citenamefont
  {Meerholz}, \citenamefont {van Loosdrecht}, \citenamefont {Lindfors},\ and\
  \citenamefont {Grüneis}}]{senkovskiy_making_2017}%
  \BibitemOpen
  \bibfield  {author} {\bibinfo {author} {\bibfnamefont {B.~V.}\ \bibnamefont
  {Senkovskiy}}, \bibinfo {author} {\bibfnamefont {M.}~\bibnamefont
  {Pfeiffer}}, \bibinfo {author} {\bibfnamefont {S.~K.}\ \bibnamefont {Alavi}},
  \bibinfo {author} {\bibfnamefont {A.}~\bibnamefont {Bliesener}}, \bibinfo
  {author} {\bibfnamefont {J.}~\bibnamefont {Zhu}}, \bibinfo {author}
  {\bibfnamefont {S.}~\bibnamefont {Michel}}, \bibinfo {author} {\bibfnamefont
  {A.~V.}\ \bibnamefont {Fedorov}}, \bibinfo {author} {\bibfnamefont
  {R.}~\bibnamefont {German}}, \bibinfo {author} {\bibfnamefont
  {D.}~\bibnamefont {Hertel}}, \bibinfo {author} {\bibfnamefont
  {D.}~\bibnamefont {Haberer}}, \bibinfo {author} {\bibfnamefont
  {L.}~\bibnamefont {Petaccia}}, \bibinfo {author} {\bibfnamefont {F.~R.}\
  \bibnamefont {Fischer}}, \bibinfo {author} {\bibfnamefont {K.}~\bibnamefont
  {Meerholz}}, \bibinfo {author} {\bibfnamefont {P.~H.~M.}\ \bibnamefont {van
  Loosdrecht}}, \bibinfo {author} {\bibfnamefont {K.}~\bibnamefont {Lindfors}},
  \ and\ \bibinfo {author} {\bibfnamefont {A.}~\bibnamefont {Grüneis}},\
  }\href {\doibase 10.1021/acs.nanolett.7b00147} {\bibfield  {journal}
  {\bibinfo  {journal} {Nano Lett.}\ }\textbf {\bibinfo {volume} {17}},\
  \bibinfo {pages} {4029} (\bibinfo {year} {2017})}\BibitemShut {NoStop}%
\bibitem [{\citenamefont {Chong}\ \emph {et~al.}(2018)\citenamefont {Chong},
  \citenamefont {Afshar-Imani}, \citenamefont {Scheurer}, \citenamefont
  {Cardoso}, \citenamefont {Ferretti}, \citenamefont {Prezzi},\ and\
  \citenamefont {Schull}}]{Chong2018}%
  \BibitemOpen
  \bibfield  {author} {\bibinfo {author} {\bibfnamefont {M.~C.}\ \bibnamefont
  {Chong}}, \bibinfo {author} {\bibfnamefont {N.}~\bibnamefont {Afshar-Imani}},
  \bibinfo {author} {\bibfnamefont {F.}~\bibnamefont {Scheurer}}, \bibinfo
  {author} {\bibfnamefont {C.}~\bibnamefont {Cardoso}}, \bibinfo {author}
  {\bibfnamefont {A.}~\bibnamefont {Ferretti}}, \bibinfo {author}
  {\bibfnamefont {D.}~\bibnamefont {Prezzi}}, \ and\ \bibinfo {author}
  {\bibfnamefont {G.}~\bibnamefont {Schull}},\ }\href {\doibase
  10.1021/acs.nanolett.7b03797} {\bibfield  {journal} {\bibinfo  {journal}
  {Nano Lett.}\ }\textbf {\bibinfo {volume} {18}},\ \bibinfo {pages} {175}
  (\bibinfo {year} {2018})}\BibitemShut {NoStop}%
\bibitem [{\citenamefont {Friedrich}\ \emph {et~al.}(2020)\citenamefont
  {Friedrich}, \citenamefont {Brandimarte}, \citenamefont {Li}, \citenamefont
  {Saito}, \citenamefont {Yamaguchi}, \citenamefont {Pozo}, \citenamefont
  {Peña}, \citenamefont {Frederiksen}, \citenamefont {Garcia-Lekue},
  \citenamefont {Sánchez-Portal},\ and\ \citenamefont
  {Pascual}}]{friedrich_magnetism_2020}%
  \BibitemOpen
  \bibfield  {author} {\bibinfo {author} {\bibfnamefont {N.}~\bibnamefont
  {Friedrich}}, \bibinfo {author} {\bibfnamefont {P.}~\bibnamefont
  {Brandimarte}}, \bibinfo {author} {\bibfnamefont {J.}~\bibnamefont {Li}},
  \bibinfo {author} {\bibfnamefont {S.}~\bibnamefont {Saito}}, \bibinfo
  {author} {\bibfnamefont {S.}~\bibnamefont {Yamaguchi}}, \bibinfo {author}
  {\bibfnamefont {I.}~\bibnamefont {Pozo}}, \bibinfo {author} {\bibfnamefont
  {D.}~\bibnamefont {Peña}}, \bibinfo {author} {\bibfnamefont
  {T.}~\bibnamefont {Frederiksen}}, \bibinfo {author} {\bibfnamefont
  {A.}~\bibnamefont {Garcia-Lekue}}, \bibinfo {author} {\bibfnamefont
  {D.}~\bibnamefont {Sánchez-Portal}}, \ and\ \bibinfo {author} {\bibfnamefont
  {J.~I.}\ \bibnamefont {Pascual}},\ }\href {\doibase
  10.1103/PhysRevLett.125.146801} {\bibfield  {journal} {\bibinfo  {journal}
  {Phys. Rev. Lett.}\ }\textbf {\bibinfo {volume} {125}},\ \bibinfo {pages}
  {146801} (\bibinfo {year} {2020})}\BibitemShut {NoStop}%
\bibitem [{\citenamefont {Carbonell-Sanromà}\ \emph
  {et~al.}(2017{\natexlab{b}})\citenamefont {Carbonell-Sanromà}, \citenamefont
  {Hieulle}, \citenamefont {Vilas-Varela}, \citenamefont {Brandimarte},
  \citenamefont {Iraola}, \citenamefont {Barragán}, \citenamefont {Li},
  \citenamefont {Abadia}, \citenamefont {Corso}, \citenamefont
  {Sánchez-Portal}, \citenamefont {Peña},\ and\ \citenamefont
  {Pascual}}]{Carbonell-Sanroma2017_edge}%
  \BibitemOpen
  \bibfield  {author} {\bibinfo {author} {\bibfnamefont {E.}~\bibnamefont
  {Carbonell-Sanromà}}, \bibinfo {author} {\bibfnamefont {J.}~\bibnamefont
  {Hieulle}}, \bibinfo {author} {\bibfnamefont {M.}~\bibnamefont
  {Vilas-Varela}}, \bibinfo {author} {\bibfnamefont {P.}~\bibnamefont
  {Brandimarte}}, \bibinfo {author} {\bibfnamefont {M.}~\bibnamefont {Iraola}},
  \bibinfo {author} {\bibfnamefont {A.}~\bibnamefont {Barragán}}, \bibinfo
  {author} {\bibfnamefont {J.}~\bibnamefont {Li}}, \bibinfo {author}
  {\bibfnamefont {M.}~\bibnamefont {Abadia}}, \bibinfo {author} {\bibfnamefont
  {M.}~\bibnamefont {Corso}}, \bibinfo {author} {\bibfnamefont
  {D.}~\bibnamefont {Sánchez-Portal}}, \bibinfo {author} {\bibfnamefont
  {D.}~\bibnamefont {Peña}}, \ and\ \bibinfo {author} {\bibfnamefont {J.~I.}\
  \bibnamefont {Pascual}},\ }\href {\doibase 10.1021/acsnano.7b03522}
  {\bibfield  {journal} {\bibinfo  {journal} {ACS Nano}\ }\textbf {\bibinfo
  {volume} {11}},\ \bibinfo {pages} {7355} (\bibinfo {year}
  {2017}{\natexlab{b}})}\BibitemShut {NoStop}%
\bibitem [{\citenamefont {Pei}\ \emph {et~al.}(2022)\citenamefont {Pei},
  \citenamefont {Thomas}, \citenamefont {Sopp}, \citenamefont {Tsang},
  \citenamefont {Dotti}, \citenamefont {Baugh}, \citenamefont {Chilton},
  \citenamefont {Cardona-Serra}, \citenamefont {Gaita-Ariño}, \citenamefont
  {Anderson},\ and\ \citenamefont {Bogani}}]{pei_exchange-induced_2022}%
  \BibitemOpen
  \bibfield  {author} {\bibinfo {author} {\bibfnamefont {T.}~\bibnamefont
  {Pei}}, \bibinfo {author} {\bibfnamefont {J.~O.}\ \bibnamefont {Thomas}},
  \bibinfo {author} {\bibfnamefont {S.}~\bibnamefont {Sopp}}, \bibinfo {author}
  {\bibfnamefont {M.-Y.}\ \bibnamefont {Tsang}}, \bibinfo {author}
  {\bibfnamefont {N.}~\bibnamefont {Dotti}}, \bibinfo {author} {\bibfnamefont
  {J.}~\bibnamefont {Baugh}}, \bibinfo {author} {\bibfnamefont {N.~F.}\
  \bibnamefont {Chilton}}, \bibinfo {author} {\bibfnamefont {S.}~\bibnamefont
  {Cardona-Serra}}, \bibinfo {author} {\bibfnamefont {A.}~\bibnamefont
  {Gaita-Ariño}}, \bibinfo {author} {\bibfnamefont {H.~L.}\ \bibnamefont
  {Anderson}}, \ and\ \bibinfo {author} {\bibfnamefont {L.}~\bibnamefont
  {Bogani}},\ }\href {\doibase 10.1038/s41467-022-31909-w} {\bibfield
  {journal} {\bibinfo  {journal} {Nature Communications}\ }\textbf {\bibinfo
  {volume} {13}},\ \bibinfo {pages} {4506} (\bibinfo {year}
  {2022})}\BibitemShut {NoStop}%
\bibitem [{\citenamefont {Niu}\ \emph {et~al.}(2023)\citenamefont {Niu},
  \citenamefont {Sopp}, \citenamefont {Lodi}, \citenamefont {Gee},
  \citenamefont {Kong}, \citenamefont {Pei}, \citenamefont {Gehring},
  \citenamefont {Nägele}, \citenamefont {Lau}, \citenamefont {Ma},
  \citenamefont {Liu}, \citenamefont {Narita}, \citenamefont {Mol},
  \citenamefont {Burghard}, \citenamefont {Müllen}, \citenamefont {Mai},
  \citenamefont {Feng},\ and\ \citenamefont {Bogani}}]{niu_exceptionally_2023}%
  \BibitemOpen
  \bibfield  {author} {\bibinfo {author} {\bibfnamefont {W.}~\bibnamefont
  {Niu}}, \bibinfo {author} {\bibfnamefont {S.}~\bibnamefont {Sopp}}, \bibinfo
  {author} {\bibfnamefont {A.}~\bibnamefont {Lodi}}, \bibinfo {author}
  {\bibfnamefont {A.}~\bibnamefont {Gee}}, \bibinfo {author} {\bibfnamefont
  {F.}~\bibnamefont {Kong}}, \bibinfo {author} {\bibfnamefont {T.}~\bibnamefont
  {Pei}}, \bibinfo {author} {\bibfnamefont {P.}~\bibnamefont {Gehring}},
  \bibinfo {author} {\bibfnamefont {J.}~\bibnamefont {Nägele}}, \bibinfo
  {author} {\bibfnamefont {C.~S.}\ \bibnamefont {Lau}}, \bibinfo {author}
  {\bibfnamefont {J.}~\bibnamefont {Ma}}, \bibinfo {author} {\bibfnamefont
  {J.}~\bibnamefont {Liu}}, \bibinfo {author} {\bibfnamefont {A.}~\bibnamefont
  {Narita}}, \bibinfo {author} {\bibfnamefont {J.}~\bibnamefont {Mol}},
  \bibinfo {author} {\bibfnamefont {M.}~\bibnamefont {Burghard}}, \bibinfo
  {author} {\bibfnamefont {K.}~\bibnamefont {Müllen}}, \bibinfo {author}
  {\bibfnamefont {Y.}~\bibnamefont {Mai}}, \bibinfo {author} {\bibfnamefont
  {X.}~\bibnamefont {Feng}}, \ and\ \bibinfo {author} {\bibfnamefont
  {L.}~\bibnamefont {Bogani}},\ }\href {\doibase 10.1038/s41563-022-01460-6}
  {\bibfield  {journal} {\bibinfo  {journal} {Nature Materials}\ }\textbf
  {\bibinfo {volume} {22}},\ \bibinfo {pages} {180} (\bibinfo {year}
  {2023})}\BibitemShut {NoStop}%
\bibitem [{\citenamefont {Chen}\ \emph {et~al.}(2023)\citenamefont {Chen},
  \citenamefont {Deng}, \citenamefont {Hou}, \citenamefont {Bian},
  \citenamefont {Swett}, \citenamefont {Wu}, \citenamefont {Baugh},
  \citenamefont {Bogani}, \citenamefont {Briggs}, \citenamefont {Mol},
  \citenamefont {Lambert}, \citenamefont {Anderson},\ and\ \citenamefont
  {Thomas}}]{chen_phase-coherent_2023}%
  \BibitemOpen
  \bibfield  {author} {\bibinfo {author} {\bibfnamefont {Z.}~\bibnamefont
  {Chen}}, \bibinfo {author} {\bibfnamefont {J.-R.}\ \bibnamefont {Deng}},
  \bibinfo {author} {\bibfnamefont {S.}~\bibnamefont {Hou}}, \bibinfo {author}
  {\bibfnamefont {X.}~\bibnamefont {Bian}}, \bibinfo {author} {\bibfnamefont
  {J.~L.}\ \bibnamefont {Swett}}, \bibinfo {author} {\bibfnamefont
  {Q.}~\bibnamefont {Wu}}, \bibinfo {author} {\bibfnamefont {J.}~\bibnamefont
  {Baugh}}, \bibinfo {author} {\bibfnamefont {L.}~\bibnamefont {Bogani}},
  \bibinfo {author} {\bibfnamefont {G.~A.~D.}\ \bibnamefont {Briggs}}, \bibinfo
  {author} {\bibfnamefont {J.~A.}\ \bibnamefont {Mol}}, \bibinfo {author}
  {\bibfnamefont {C.~J.}\ \bibnamefont {Lambert}}, \bibinfo {author}
  {\bibfnamefont {H.~L.}\ \bibnamefont {Anderson}}, \ and\ \bibinfo {author}
  {\bibfnamefont {J.~O.}\ \bibnamefont {Thomas}},\ }\href {\doibase
  10.1021/jacs.3c02451} {\bibfield  {journal} {\bibinfo  {journal} {Journal of
  the American Chemical Society}\ }\textbf {\bibinfo {volume} {145}},\ \bibinfo
  {pages} {15265} (\bibinfo {year} {2023})}\BibitemShut {NoStop}%
\bibitem [{\citenamefont {Zhang}\ \emph {et~al.}(2023)\citenamefont {Zhang},
  \citenamefont {Qian}, \citenamefont {Barin}, \citenamefont {Daaoub},
  \citenamefont {Chen}, \citenamefont {Müllen}, \citenamefont {Sangtarash},
  \citenamefont {Ruffieux}, \citenamefont {Fasel}, \citenamefont {Sadeghi},
  \citenamefont {Zhang}, \citenamefont {Calame},\ and\ \citenamefont
  {Perrin}}]{Zhang2023}%
  \BibitemOpen
  \bibfield  {author} {\bibinfo {author} {\bibfnamefont {J.}~\bibnamefont
  {Zhang}}, \bibinfo {author} {\bibfnamefont {L.}~\bibnamefont {Qian}},
  \bibinfo {author} {\bibfnamefont {G.~B.}\ \bibnamefont {Barin}}, \bibinfo
  {author} {\bibfnamefont {A.~H.~S.}\ \bibnamefont {Daaoub}}, \bibinfo {author}
  {\bibfnamefont {P.}~\bibnamefont {Chen}}, \bibinfo {author} {\bibfnamefont
  {K.}~\bibnamefont {Müllen}}, \bibinfo {author} {\bibfnamefont
  {S.}~\bibnamefont {Sangtarash}}, \bibinfo {author} {\bibfnamefont
  {P.}~\bibnamefont {Ruffieux}}, \bibinfo {author} {\bibfnamefont
  {R.}~\bibnamefont {Fasel}}, \bibinfo {author} {\bibfnamefont
  {H.}~\bibnamefont {Sadeghi}}, \bibinfo {author} {\bibfnamefont
  {J.}~\bibnamefont {Zhang}}, \bibinfo {author} {\bibfnamefont
  {M.}~\bibnamefont {Calame}}, \ and\ \bibinfo {author} {\bibfnamefont {M.~L.}\
  \bibnamefont {Perrin}},\ }\href {\doibase
  https://doi.org/10.1038/s41928-023-00991-3} {\bibfield  {journal} {\bibinfo
  {journal} {Nature electronics}\ }\textbf {\bibinfo {volume} {6}},\ \bibinfo
  {pages} {572} (\bibinfo {year} {2023})}\BibitemShut {NoStop}%
\bibitem [{\citenamefont {Lafferentz}\ \emph {et~al.}(2009)\citenamefont
  {Lafferentz}, \citenamefont {Ample}, \citenamefont {Yu}, \citenamefont
  {Hecht}, \citenamefont {Joachim},\ and\ \citenamefont
  {Grill}}]{Lafferentz2009}%
  \BibitemOpen
  \bibfield  {author} {\bibinfo {author} {\bibfnamefont {L.}~\bibnamefont
  {Lafferentz}}, \bibinfo {author} {\bibfnamefont {F.}~\bibnamefont {Ample}},
  \bibinfo {author} {\bibfnamefont {H.}~\bibnamefont {Yu}}, \bibinfo {author}
  {\bibfnamefont {S.}~\bibnamefont {Hecht}}, \bibinfo {author} {\bibfnamefont
  {C.}~\bibnamefont {Joachim}}, \ and\ \bibinfo {author} {\bibfnamefont
  {L.}~\bibnamefont {Grill}},\ }\href {\doibase 10.1126/science.1168255}
  {\bibfield  {journal} {\bibinfo  {journal} {Science}\ }\textbf {\bibinfo
  {volume} {323}},\ \bibinfo {pages} {1193} (\bibinfo {year}
  {2009})}\BibitemShut {NoStop}%
\bibitem [{\citenamefont {Koch}\ \emph {et~al.}(2012)\citenamefont {Koch},
  \citenamefont {Ample}, \citenamefont {Joachim},\ and\ \citenamefont
  {Grill}}]{Koch2012}%
  \BibitemOpen
  \bibfield  {author} {\bibinfo {author} {\bibfnamefont {M.}~\bibnamefont
  {Koch}}, \bibinfo {author} {\bibfnamefont {F.}~\bibnamefont {Ample}},
  \bibinfo {author} {\bibfnamefont {C.}~\bibnamefont {Joachim}}, \ and\
  \bibinfo {author} {\bibfnamefont {L.}~\bibnamefont {Grill}},\ }\href
  {\doibase 10.1038/nnano.2012.169} {\bibfield  {journal} {\bibinfo  {journal}
  {Nat. Nanotechnol.}\ }\textbf {\bibinfo {volume} {7}},\ \bibinfo {pages}
  {713} (\bibinfo {year} {2012})}\BibitemShut {NoStop}%
\bibitem [{\citenamefont {Jacobse}\ \emph {et~al.}(2018)\citenamefont
  {Jacobse}, \citenamefont {Mangnus}, \citenamefont {Zevenhuizen},\ and\
  \citenamefont {Swart}}]{Jacobse2018}%
  \BibitemOpen
  \bibfield  {author} {\bibinfo {author} {\bibfnamefont {P.~H.}\ \bibnamefont
  {Jacobse}}, \bibinfo {author} {\bibfnamefont {M.~J.~J.}\ \bibnamefont
  {Mangnus}}, \bibinfo {author} {\bibfnamefont {S.~J.~M.}\ \bibnamefont
  {Zevenhuizen}}, \ and\ \bibinfo {author} {\bibfnamefont {I.}~\bibnamefont
  {Swart}},\ }\href {\doibase 10.1021/acsnano.8b02770} {\bibfield  {journal}
  {\bibinfo  {journal} {ACS Nano}\ }\textbf {\bibinfo {volume} {12}},\ \bibinfo
  {pages} {7048} (\bibinfo {year} {2018})}\BibitemShut {NoStop}%
\bibitem [{\citenamefont {Li}\ \emph {et~al.}(2019)\citenamefont {Li},
  \citenamefont {Friedrich}, \citenamefont {Merino}, \citenamefont {de~Oteyza},
  \citenamefont {Peña}, \citenamefont {Jacob},\ and\ \citenamefont
  {Pascual}}]{Li2019}%
  \BibitemOpen
  \bibfield  {author} {\bibinfo {author} {\bibfnamefont {J.}~\bibnamefont
  {Li}}, \bibinfo {author} {\bibfnamefont {N.}~\bibnamefont {Friedrich}},
  \bibinfo {author} {\bibfnamefont {N.}~\bibnamefont {Merino}}, \bibinfo
  {author} {\bibfnamefont {D.~G.}\ \bibnamefont {de~Oteyza}}, \bibinfo {author}
  {\bibfnamefont {D.}~\bibnamefont {Peña}}, \bibinfo {author} {\bibfnamefont
  {D.}~\bibnamefont {Jacob}}, \ and\ \bibinfo {author} {\bibfnamefont {J.~I.}\
  \bibnamefont {Pascual}},\ }\href {\doibase 10.1021/acs.nanolett.9b00883}
  {\bibfield  {journal} {\bibinfo  {journal} {Nano Lett.}\ }\textbf {\bibinfo
  {volume} {19}},\ \bibinfo {pages} {3288} (\bibinfo {year}
  {2019})}\BibitemShut {NoStop}%
\bibitem [{\citenamefont {Lawrence}\ \emph {et~al.}(2020)\citenamefont
  {Lawrence}, \citenamefont {Brandimarte}, \citenamefont {Berdonces-Layunta},
  \citenamefont {Mohammed}, \citenamefont {Grewal}, \citenamefont {Leon},
  \citenamefont {Sánchez-Portal},\ and\ \citenamefont
  {de~Oteyza}}]{lawrence_probing_2020}%
  \BibitemOpen
  \bibfield  {author} {\bibinfo {author} {\bibfnamefont {J.}~\bibnamefont
  {Lawrence}}, \bibinfo {author} {\bibfnamefont {P.}~\bibnamefont
  {Brandimarte}}, \bibinfo {author} {\bibfnamefont {A.}~\bibnamefont
  {Berdonces-Layunta}}, \bibinfo {author} {\bibfnamefont {M.~S.~G.}\
  \bibnamefont {Mohammed}}, \bibinfo {author} {\bibfnamefont {A.}~\bibnamefont
  {Grewal}}, \bibinfo {author} {\bibfnamefont {C.~C.}\ \bibnamefont {Leon}},
  \bibinfo {author} {\bibfnamefont {D.}~\bibnamefont {Sánchez-Portal}}, \ and\
  \bibinfo {author} {\bibfnamefont {D.~G.}\ \bibnamefont {de~Oteyza}},\ }\href
  {\doibase 10.1021/acsnano.9b10191} {\bibfield  {journal} {\bibinfo  {journal}
  {ACS Nano}\ }\textbf {\bibinfo {volume} {14}},\ \bibinfo {pages} {4499}
  (\bibinfo {year} {2020})}\BibitemShut {NoStop}%
\bibitem [{\citenamefont {Jiang}\ \emph {et~al.}(2022)\citenamefont {Jiang},
  \citenamefont {Scheurer}, \citenamefont {Sun}, \citenamefont {Ruffieux},
  \citenamefont {Yao}, \citenamefont {Narita}, \citenamefont {Mullen},
  \citenamefont {Fasel}, \citenamefont {Frederiksen},\ and\ \citenamefont
  {Schull}}]{jiang_length-independent_2022}%
  \BibitemOpen
  \bibfield  {author} {\bibinfo {author} {\bibfnamefont {S.}~\bibnamefont
  {Jiang}}, \bibinfo {author} {\bibfnamefont {F.}~\bibnamefont {Scheurer}},
  \bibinfo {author} {\bibfnamefont {Q.}~\bibnamefont {Sun}}, \bibinfo {author}
  {\bibfnamefont {P.}~\bibnamefont {Ruffieux}}, \bibinfo {author}
  {\bibfnamefont {X.}~\bibnamefont {Yao}}, \bibinfo {author} {\bibfnamefont
  {A.}~\bibnamefont {Narita}}, \bibinfo {author} {\bibfnamefont
  {K.}~\bibnamefont {Mullen}}, \bibinfo {author} {\bibfnamefont
  {R.}~\bibnamefont {Fasel}}, \bibinfo {author} {\bibfnamefont
  {T.}~\bibnamefont {Frederiksen}}, \ and\ \bibinfo {author} {\bibfnamefont
  {G.}~\bibnamefont {Schull}},\ }\href {\doibase
  https://doi.org/10.48550/arXiv.2208.03145} {\enquote {\bibinfo {title}
  {{Length-independent quantum transport through topological band states of
  graphene nanoribbons}},}\ } (\bibinfo {year} {2022})\BibitemShut {NoStop}%
\bibitem [{\citenamefont {Zhang}\ \emph {et~al.}(2022)\citenamefont {Zhang},
  \citenamefont {Li}, \citenamefont {Dong}, \citenamefont {Zhu}, \citenamefont
  {Zheng},\ and\ \citenamefont {Zhang}}]{Zhang2022}%
  \BibitemOpen
  \bibfield  {author} {\bibinfo {author} {\bibfnamefont {P.}~\bibnamefont
  {Zhang}}, \bibinfo {author} {\bibfnamefont {X.}~\bibnamefont {Li}}, \bibinfo
  {author} {\bibfnamefont {J.}~\bibnamefont {Dong}}, \bibinfo {author}
  {\bibfnamefont {M.}~\bibnamefont {Zhu}}, \bibinfo {author} {\bibfnamefont
  {F.}~\bibnamefont {Zheng}}, \ and\ \bibinfo {author} {\bibfnamefont
  {J.}~\bibnamefont {Zhang}},\ }\href {\doibase 10.1063/5.0086377} {\bibfield
  {journal} {\bibinfo  {journal} {Appl. Phys. Lett.}\ }\textbf {\bibinfo
  {volume} {120}},\ \bibinfo {pages} {132406} (\bibinfo {year}
  {2022})}\BibitemShut {NoStop}%
\bibitem [{\citenamefont {Qiu}\ \emph {et~al.}(2004)\citenamefont {Qiu},
  \citenamefont {Nazin},\ and\ \citenamefont {Ho}}]{qiu_vibronic_2004}%
  \BibitemOpen
  \bibfield  {author} {\bibinfo {author} {\bibfnamefont {X.~H.}\ \bibnamefont
  {Qiu}}, \bibinfo {author} {\bibfnamefont {G.~V.}\ \bibnamefont {Nazin}}, \
  and\ \bibinfo {author} {\bibfnamefont {W.}~\bibnamefont {Ho}},\ }\href
  {\doibase 10.1103/PhysRevLett.92.206102} {\bibfield  {journal} {\bibinfo
  {journal} {Phys. Rev. Lett.}\ }\textbf {\bibinfo {volume} {92}},\ \bibinfo
  {pages} {206102} (\bibinfo {year} {2004})}\BibitemShut {NoStop}%
\bibitem [{\citenamefont {Nazin}\ \emph {et~al.}(2005)\citenamefont {Nazin},
  \citenamefont {Wu},\ and\ \citenamefont {Ho}}]{nazin_tunneling_2005}%
  \BibitemOpen
  \bibfield  {author} {\bibinfo {author} {\bibfnamefont {G.~V.}\ \bibnamefont
  {Nazin}}, \bibinfo {author} {\bibfnamefont {S.~W.}\ \bibnamefont {Wu}}, \
  and\ \bibinfo {author} {\bibfnamefont {W.}~\bibnamefont {Ho}},\ }\href
  {\doibase 10.1073/pnas.0501171102} {\bibfield  {journal} {\bibinfo  {journal}
  {PNAS}\ }\textbf {\bibinfo {volume} {102}},\ \bibinfo {pages} {8832}
  (\bibinfo {year} {2005})}\BibitemShut {NoStop}%
\bibitem [{\citenamefont {Wagner}\ \emph {et~al.}(2015)\citenamefont {Wagner},
  \citenamefont {Green}, \citenamefont {Leinen}, \citenamefont {Deilmann},
  \citenamefont {Krüger}, \citenamefont {Rohlfing}, \citenamefont {Temirov},\
  and\ \citenamefont {Tautz}}]{Wagner2015sqdm}%
  \BibitemOpen
  \bibfield  {author} {\bibinfo {author} {\bibfnamefont {C.}~\bibnamefont
  {Wagner}}, \bibinfo {author} {\bibfnamefont {M.~F.}\ \bibnamefont {Green}},
  \bibinfo {author} {\bibfnamefont {P.}~\bibnamefont {Leinen}}, \bibinfo
  {author} {\bibfnamefont {T.}~\bibnamefont {Deilmann}}, \bibinfo {author}
  {\bibfnamefont {P.}~\bibnamefont {Krüger}}, \bibinfo {author} {\bibfnamefont
  {M.}~\bibnamefont {Rohlfing}}, \bibinfo {author} {\bibfnamefont
  {R.}~\bibnamefont {Temirov}}, \ and\ \bibinfo {author} {\bibfnamefont
  {F.~S.}\ \bibnamefont {Tautz}},\ }\href {\doibase
  10.1103/PhysRevLett.115.026101} {\bibfield  {journal} {\bibinfo  {journal}
  {Phys. Rev. Lett.}\ }\textbf {\bibinfo {volume} {115}},\ \bibinfo {pages} {1}
  (\bibinfo {year} {2015})}\BibitemShut {NoStop}%
\bibitem [{\citenamefont {Merino-Díez}\ \emph {et~al.}(2018)\citenamefont
  {Merino-Díez}, \citenamefont {Li}, \citenamefont {Garcia-Lekue},
  \citenamefont {Vasseur}, \citenamefont {Vilas-Varela}, \citenamefont
  {Carbonell-Sanromà}, \citenamefont {Corso}, \citenamefont {Ortega},
  \citenamefont {Peña}, \citenamefont {Pascual},\ and\ \citenamefont
  {de~Oteyza}}]{Merino-Diez2018}%
  \BibitemOpen
  \bibfield  {author} {\bibinfo {author} {\bibfnamefont {N.}~\bibnamefont
  {Merino-Díez}}, \bibinfo {author} {\bibfnamefont {J.}~\bibnamefont {Li}},
  \bibinfo {author} {\bibfnamefont {A.}~\bibnamefont {Garcia-Lekue}}, \bibinfo
  {author} {\bibfnamefont {G.}~\bibnamefont {Vasseur}}, \bibinfo {author}
  {\bibfnamefont {M.}~\bibnamefont {Vilas-Varela}}, \bibinfo {author}
  {\bibfnamefont {E.}~\bibnamefont {Carbonell-Sanromà}}, \bibinfo {author}
  {\bibfnamefont {M.}~\bibnamefont {Corso}}, \bibinfo {author} {\bibfnamefont
  {J.~E.}\ \bibnamefont {Ortega}}, \bibinfo {author} {\bibfnamefont
  {D.}~\bibnamefont {Peña}}, \bibinfo {author} {\bibfnamefont {J.~I.}\
  \bibnamefont {Pascual}}, \ and\ \bibinfo {author} {\bibfnamefont {D.~G.}\
  \bibnamefont {de~Oteyza}},\ }\href {\doibase 10.1021/acs.jpclett.7b02767}
  {\bibfield  {journal} {\bibinfo  {journal} {J. Phys. Chem. Lett.}\ }\textbf
  {\bibinfo {volume} {9}},\ \bibinfo {pages} {25} (\bibinfo {year}
  {2018})}\BibitemShut {NoStop}%
\bibitem [{\citenamefont {Perrin}\ \emph {et~al.}(2016)\citenamefont {Perrin},
  \citenamefont {Gal{\'a}n}, \citenamefont {Eelkema}, \citenamefont {Thijssen},
  \citenamefont {Grozema},\ and\ \citenamefont {van~der
  Zant}}]{perrin_gate-tunable_2016}%
  \BibitemOpen
  \bibfield  {author} {\bibinfo {author} {\bibfnamefont {M.~L.}\ \bibnamefont
  {Perrin}}, \bibinfo {author} {\bibfnamefont {E.}~\bibnamefont {Gal{\'a}n}},
  \bibinfo {author} {\bibfnamefont {R.}~\bibnamefont {Eelkema}}, \bibinfo
  {author} {\bibfnamefont {J.~M.}\ \bibnamefont {Thijssen}}, \bibinfo {author}
  {\bibfnamefont {F.}~\bibnamefont {Grozema}}, \ and\ \bibinfo {author}
  {\bibfnamefont {H.~S.}\ \bibnamefont {van~der Zant}},\ }\href {\doibase
  10.1039/C6NR00735J} {\bibfield  {journal} {\bibinfo  {journal} {Nanoscale}\
  }\textbf {\bibinfo {volume} {8}},\ \bibinfo {pages} {8919} (\bibinfo {year}
  {2016})}\BibitemShut {NoStop}%
\bibitem [{\citenamefont {Franke}\ and\ \citenamefont
  {Pascual}(2012)}]{franke_effects_2012}%
  \BibitemOpen
  \bibfield  {author} {\bibinfo {author} {\bibfnamefont {K.~J.}\ \bibnamefont
  {Franke}}\ and\ \bibinfo {author} {\bibfnamefont {J.~I.}\ \bibnamefont
  {Pascual}},\ }\href {\doibase 10.1088/0953-8984/24/39/394002} {\bibfield
  {journal} {\bibinfo  {journal} {J. Phys.: Condens. Matter}\ }\textbf
  {\bibinfo {volume} {24}},\ \bibinfo {pages} {394002} (\bibinfo {year}
  {2012})}\BibitemShut {NoStop}%
\bibitem [{\citenamefont {Chen}\ \emph
  {et~al.}(2017{\natexlab{b}})\citenamefont {Chen}, \citenamefont {Roemer},
  \citenamefont {Yuan}, \citenamefont {Du}, \citenamefont {Thompson},
  \citenamefont {del Barco},\ and\ \citenamefont
  {Nijhuis}}]{chen_molecular_2017}%
  \BibitemOpen
  \bibfield  {author} {\bibinfo {author} {\bibfnamefont {X.}~\bibnamefont
  {Chen}}, \bibinfo {author} {\bibfnamefont {M.}~\bibnamefont {Roemer}},
  \bibinfo {author} {\bibfnamefont {L.}~\bibnamefont {Yuan}}, \bibinfo {author}
  {\bibfnamefont {W.}~\bibnamefont {Du}}, \bibinfo {author} {\bibfnamefont
  {D.}~\bibnamefont {Thompson}}, \bibinfo {author} {\bibfnamefont
  {E.}~\bibnamefont {del Barco}}, \ and\ \bibinfo {author} {\bibfnamefont
  {C.~A.}\ \bibnamefont {Nijhuis}},\ }\href {\doibase 10.1038/nnano.2017.110}
  {\bibfield  {journal} {\bibinfo  {journal} {Nat. Nanotechnol.}\ }\textbf
  {\bibinfo {volume} {12}},\ \bibinfo {pages} {797} (\bibinfo {year}
  {2017}{\natexlab{b}})}\BibitemShut {NoStop}%
\bibitem [{\citenamefont {Zhang}\ \emph {et~al.}(2021)\citenamefont {Zhang},
  \citenamefont {Chen}, \citenamefont {Zhao}, \citenamefont {Sun},
  \citenamefont {Shi}, \citenamefont {Wang}, \citenamefont {Hu},\ and\
  \citenamefont {Wang}}]{zhang_large_2021}%
  \BibitemOpen
  \bibfield  {author} {\bibinfo {author} {\bibfnamefont {G.-P.}\ \bibnamefont
  {Zhang}}, \bibinfo {author} {\bibfnamefont {L.-Y.}\ \bibnamefont {Chen}},
  \bibinfo {author} {\bibfnamefont {J.-M.}\ \bibnamefont {Zhao}}, \bibinfo
  {author} {\bibfnamefont {Y.-Z.}\ \bibnamefont {Sun}}, \bibinfo {author}
  {\bibfnamefont {N.-P.}\ \bibnamefont {Shi}}, \bibinfo {author} {\bibfnamefont
  {M.-L.}\ \bibnamefont {Wang}}, \bibinfo {author} {\bibfnamefont {G.-C.}\
  \bibnamefont {Hu}}, \ and\ \bibinfo {author} {\bibfnamefont {C.-K.}\
  \bibnamefont {Wang}},\ }\href {\doibase 10.1021/acs.jpcc.1c04093} {\bibfield
  {journal} {\bibinfo  {journal} {J. Phys. Chem. C}\ }\textbf {\bibinfo
  {volume} {125}},\ \bibinfo {pages} {20783} (\bibinfo {year}
  {2021})}\BibitemShut {NoStop}%
\bibitem [{\citenamefont {Söde}\ \emph {et~al.}(2015)\citenamefont {Söde},
  \citenamefont {Talirz}, \citenamefont {Gröning}, \citenamefont {Pignedoli},
  \citenamefont {Berger}, \citenamefont {Feng}, \citenamefont {Müllen},
  \citenamefont {Fasel},\ and\ \citenamefont {Ruffieux}}]{Sode2015}%
  \BibitemOpen
  \bibfield  {author} {\bibinfo {author} {\bibfnamefont {H.}~\bibnamefont
  {Söde}}, \bibinfo {author} {\bibfnamefont {L.}~\bibnamefont {Talirz}},
  \bibinfo {author} {\bibfnamefont {O.}~\bibnamefont {Gröning}}, \bibinfo
  {author} {\bibfnamefont {C.~A.}\ \bibnamefont {Pignedoli}}, \bibinfo {author}
  {\bibfnamefont {R.}~\bibnamefont {Berger}}, \bibinfo {author} {\bibfnamefont
  {X.}~\bibnamefont {Feng}}, \bibinfo {author} {\bibfnamefont {K.}~\bibnamefont
  {Müllen}}, \bibinfo {author} {\bibfnamefont {R.}~\bibnamefont {Fasel}}, \
  and\ \bibinfo {author} {\bibfnamefont {P.}~\bibnamefont {Ruffieux}},\ }\href
  {\doibase 10.1103/PhysRevB.91.045429} {\bibfield  {journal} {\bibinfo
  {journal} {Physical Review B - Condensed Matter and Materials Physics}\
  }\textbf {\bibinfo {volume} {91}},\ \bibinfo {pages} {1} (\bibinfo {year}
  {2015})}\BibitemShut {NoStop}%
\bibitem [{\citenamefont {Elbing}\ \emph {et~al.}(2005)\citenamefont {Elbing},
  \citenamefont {Ochs}, \citenamefont {Koentopp}, \citenamefont {Fischer},
  \citenamefont {von Hänisch}, \citenamefont {Weigend}, \citenamefont {Evers},
  \citenamefont {Weber},\ and\ \citenamefont {Mayor}}]{Elbing2005}%
  \BibitemOpen
  \bibfield  {author} {\bibinfo {author} {\bibfnamefont {M.}~\bibnamefont
  {Elbing}}, \bibinfo {author} {\bibfnamefont {R.}~\bibnamefont {Ochs}},
  \bibinfo {author} {\bibfnamefont {M.}~\bibnamefont {Koentopp}}, \bibinfo
  {author} {\bibfnamefont {M.}~\bibnamefont {Fischer}}, \bibinfo {author}
  {\bibfnamefont {C.}~\bibnamefont {von Hänisch}}, \bibinfo {author}
  {\bibfnamefont {F.}~\bibnamefont {Weigend}}, \bibinfo {author} {\bibfnamefont
  {F.}~\bibnamefont {Evers}}, \bibinfo {author} {\bibfnamefont {H.~B.}\
  \bibnamefont {Weber}}, \ and\ \bibinfo {author} {\bibfnamefont
  {M.}~\bibnamefont {Mayor}},\ }\href {\doibase 10.1073/pnas.0408888102}
  {\bibfield  {journal} {\bibinfo  {journal} {PNAS}\ }\textbf {\bibinfo
  {volume} {102}},\ \bibinfo {pages} {8815} (\bibinfo {year}
  {2005})}\BibitemShut {NoStop}%
\bibitem [{\citenamefont {Perrin}\ \emph {et~al.}(2014)\citenamefont {Perrin},
  \citenamefont {Frisenda}, \citenamefont {Koole}, \citenamefont {Seldenthuis},
  \citenamefont {Gil}, \citenamefont {Valkenier}, \citenamefont {Hummelen},
  \citenamefont {Renaud}, \citenamefont {Grozema}, \citenamefont {Thijssen},
  \citenamefont {Dulić},\ and\ \citenamefont {van~der
  Zant}}]{perrin_large_2014}%
  \BibitemOpen
  \bibfield  {author} {\bibinfo {author} {\bibfnamefont {M.~L.}\ \bibnamefont
  {Perrin}}, \bibinfo {author} {\bibfnamefont {R.}~\bibnamefont {Frisenda}},
  \bibinfo {author} {\bibfnamefont {M.}~\bibnamefont {Koole}}, \bibinfo
  {author} {\bibfnamefont {J.~S.}\ \bibnamefont {Seldenthuis}}, \bibinfo
  {author} {\bibfnamefont {J.~A.~C.}\ \bibnamefont {Gil}}, \bibinfo {author}
  {\bibfnamefont {H.}~\bibnamefont {Valkenier}}, \bibinfo {author}
  {\bibfnamefont {J.~C.}\ \bibnamefont {Hummelen}}, \bibinfo {author}
  {\bibfnamefont {N.}~\bibnamefont {Renaud}}, \bibinfo {author} {\bibfnamefont
  {F.~C.}\ \bibnamefont {Grozema}}, \bibinfo {author} {\bibfnamefont {J.~M.}\
  \bibnamefont {Thijssen}}, \bibinfo {author} {\bibfnamefont {D.}~\bibnamefont
  {Dulić}}, \ and\ \bibinfo {author} {\bibfnamefont {H.~S.~J.}\ \bibnamefont
  {van~der Zant}},\ }\href {\doibase 10.1038/nnano.2014.177} {\bibfield
  {journal} {\bibinfo  {journal} {Nat. Nanotechnol.}\ }\textbf {\bibinfo
  {volume} {9}},\ \bibinfo {pages} {830} (\bibinfo {year} {2014})}\BibitemShut
  {NoStop}%
\bibitem [{\citenamefont {Carbonell-Sanromà}\ \emph
  {et~al.}(2018)\citenamefont {Carbonell-Sanromà}, \citenamefont
  {Garcia-Lekue}, \citenamefont {Corso}, \citenamefont {Vasseur}, \citenamefont
  {Brandimarte}, \citenamefont {Lobo-Checa}, \citenamefont {de~Oteyza},
  \citenamefont {Li}, \citenamefont {Kawai}, \citenamefont {Saito},
  \citenamefont {Yamaguchi}, \citenamefont {Ortega}, \citenamefont
  {Sánchez-Portal},\ and\ \citenamefont {Pascual}}]{Carbonell-Sanroma2018}%
  \BibitemOpen
  \bibfield  {author} {\bibinfo {author} {\bibfnamefont {E.}~\bibnamefont
  {Carbonell-Sanromà}}, \bibinfo {author} {\bibfnamefont {A.}~\bibnamefont
  {Garcia-Lekue}}, \bibinfo {author} {\bibfnamefont {M.}~\bibnamefont {Corso}},
  \bibinfo {author} {\bibfnamefont {G.}~\bibnamefont {Vasseur}}, \bibinfo
  {author} {\bibfnamefont {P.}~\bibnamefont {Brandimarte}}, \bibinfo {author}
  {\bibfnamefont {J.}~\bibnamefont {Lobo-Checa}}, \bibinfo {author}
  {\bibfnamefont {D.~G.}\ \bibnamefont {de~Oteyza}}, \bibinfo {author}
  {\bibfnamefont {J.}~\bibnamefont {Li}}, \bibinfo {author} {\bibfnamefont
  {S.}~\bibnamefont {Kawai}}, \bibinfo {author} {\bibfnamefont
  {S.}~\bibnamefont {Saito}}, \bibinfo {author} {\bibfnamefont
  {S.}~\bibnamefont {Yamaguchi}}, \bibinfo {author} {\bibfnamefont {J.~E.}\
  \bibnamefont {Ortega}}, \bibinfo {author} {\bibfnamefont {D.}~\bibnamefont
  {Sánchez-Portal}}, \ and\ \bibinfo {author} {\bibfnamefont {J.~I.}\
  \bibnamefont {Pascual}},\ }\href {\doibase 10.1021/acs.jpcc.8b03748}
  {\bibfield  {journal} {\bibinfo  {journal} {J. Phys. Chem. C}\ }\textbf
  {\bibinfo {volume} {122}},\ \bibinfo {pages} {16092} (\bibinfo {year}
  {2018})}\BibitemShut {NoStop}%
\bibitem [{\citenamefont {Horcas}\ \emph {et~al.}(2007)\citenamefont {Horcas},
  \citenamefont {Fernández}, \citenamefont {Gómez-Rodríguez}, \citenamefont
  {Colchero}, \citenamefont {Gómez-Herrero},\ and\ \citenamefont
  {Baro}}]{horcas_wsxm_2007}%
  \BibitemOpen
  \bibfield  {author} {\bibinfo {author} {\bibfnamefont {I.}~\bibnamefont
  {Horcas}}, \bibinfo {author} {\bibfnamefont {R.}~\bibnamefont {Fernández}},
  \bibinfo {author} {\bibfnamefont {J.}~\bibnamefont {Gómez-Rodríguez}},
  \bibinfo {author} {\bibfnamefont {J.}~\bibnamefont {Colchero}}, \bibinfo
  {author} {\bibfnamefont {J.}~\bibnamefont {Gómez-Herrero}}, \ and\ \bibinfo
  {author} {\bibfnamefont {A.}~\bibnamefont {Baro}},\ }\href {\doibase
  https://doi.org/10.1063/1.2432410} {\bibfield  {journal} {\bibinfo  {journal}
  {Rev. Sci. Instrum.}\ }\textbf {\bibinfo {volume} {78}},\ \bibinfo {pages}
  {013705} (\bibinfo {year} {2007})}\BibitemShut {NoStop}%
\bibitem [{\citenamefont {Hunter}(2007)}]{hunter_matplotlib_2007}%
  \BibitemOpen
  \bibfield  {author} {\bibinfo {author} {\bibfnamefont {J.~D.}\ \bibnamefont
  {Hunter}},\ }\href {\doibase 10.1109/MCSE.2007.55} {\bibfield  {journal}
  {\bibinfo  {journal} {Computing in Science and Engineering}\ }\textbf
  {\bibinfo {volume} {9}},\ \bibinfo {pages} {99} (\bibinfo {year}
  {2007})}\BibitemShut {NoStop}%
\bibitem [{\citenamefont {Kovesi}(2015)}]{kovesi_good_2015}%
  \BibitemOpen
  \bibfield  {author} {\bibinfo {author} {\bibfnamefont {P.}~\bibnamefont
  {Kovesi}},\ }\href {\doibase https://doi.org/10.48550/arXiv.1509.03700}
  {\enquote {\bibinfo {title} {{Good colour maps: how to design them}},}\ }
  (\bibinfo {year} {2015})\BibitemShut {NoStop}%
\end{thebibliography}
\end{document}


\title{Supplementary Information: Tunable Current Rectification through a Designer Graphene Nanoribbon}
  \author{Niklas Friedrich} \email{n.friedrich@nanogune.eu}
         \affiliation{CIC nanoGUNE-BRTA, 20018 Donostia-San Sebasti\'an, Spain}

  \author{Jingcheng Li}
         \affiliation{CIC nanoGUNE-BRTA, 20018 Donostia-San Sebasti\'an, Spain}
         \affiliation{School of Physics, Sun Yat-sen University, Guangzhou 510275, China}
 
  \author{Iago Pozo}
        \affiliation{CiQUS, Centro Singular de Investigaci\'on en Qu\'{\i}mica Biol\'oxica e Materiais Moleculares, 15705 Santiago de Compostela, Spain}
        
   \author{Diego Pe\~na} 
        \affiliation{CiQUS, Centro Singular de Investigaci\'on en Qu\'{\i}mica Biol\'oxica e Materiais Moleculares, 15705 Santiago de Compostela, Spain}
  
  \author{Jos\'e Ignacio Pascual} \email{ji.pascual@nanogune.eu}
        \affiliation{CIC nanoGUNE-BRTA, 20018 Donostia-San Sebasti\'an, Spain}
        \affiliation{Ikerbasque, Basque Foundation for Science, 48013 Bilbao, Spain}

\date{November 2023}

{
\let\clearpage\relax
\maketitle
}
\tableofcontents 

\newpage

\section{Additional discussion of $V_\text{2B}^-$ and $V_\text{2B}^+$}
\FloatBarrier

\begin{figure}[h]
    \centering
    \includegraphics{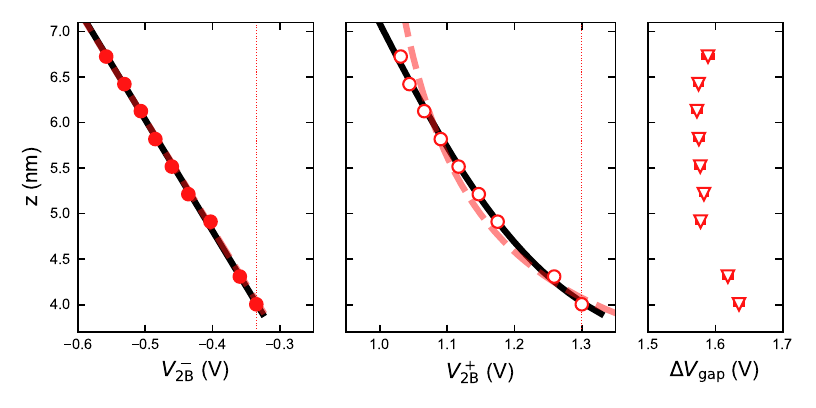}
    \caption{\label{fig:efig_V2B_z-dependence}
    \textbf{Unipolar hole transport through $V_\text{2B}^-$ and $V_\text{2B}^+$}
    Position of the resonances $V_\text{2B}^-$ and $V_\text{2B}^+$, and their separation $\Delta V_\text{gap} = V_\text{2B}^+ - V_\text{2B}^-$ as a function of $z$. The data is extracted from the spectra presented in Fig.~2\textbf{a} of the main manuscript. The red dotted lines serve as a guide to the eye. The black solid and red dashed lines are obtained by fitting the model for unipolar transport with (black) and without (red) a linear shift of the Fermi level to the experimental data. The obtained fitting parameters are $d=\SI{2.9}{nm}$, $\varepsilon_+=-\SI{0.07}{eV}$, $V_\text{sd}^+=\SI{1.3}{V}$, $E_{\text{F,}0}=-\SI{6}{meV}$ and $E_{\text{F,}z}=-\SI{60}{meV/nm}$ (black curves) and $d=\SI{2.6}{nm}$, $\varepsilon_+=-\SI{0.22}{eV}$ and $V_\text{sd}^+=\SI{0.7}{V}$ (red curves), where $E_{\text{F,}0}$ and $E_{\text{F,}z}$ describes the linear shift of the Fermi level. The decrease of $\Delta V_\text{gap}$ upon increasing $z$ is contradicting ambipolar transport \textit{via} two states in constant separation. 
	}
\end{figure}

In the following, we discuss the $z$-dependency of the resonances $V_\text{2B}^-$ and $V_\text{2B}^+$. Both $V_\text{2B}^-$ and $V_\text{2B}^+$ stay consistently between the peaks of resonant band transport (labelled $V^-$ and $V^+$) unambiguously confirming the assignment to transport \textit{via} the topological in-gap states embedded in the ribbon. Fig.~\ref{fig:efig_V2B_z-dependence} presents the position of both resonances as a function of $z$. It is evident that both peaks shift towards negative values upon increasing $z$, a first indicator against the conventional assumption of transport \textit{via} an occupied and unoccupied level at negative and positive bias voltage, respectively. Interestingly, $V_\text{2B}^+$ shifts faster to negative bias than $V_\text{2B}^-$, \textit{i.e.} the separation $\Delta V_\text{gap} = V_\text{2B}^+ - V_\text{2B}^-$ decreases (see Fig.~\ref{fig:efig_V2B_z-dependence}). The effect is especially pronounced at lower $z$. A closing of the perceived gap is inconsistent with transport enabled by two gated states (one occupied and one unoccupied) with a constant energy gap. On the other hand, a reduction of $\Delta V_\text{gap}$ is predicted in the unipolar transport model for small $z$. The slight increase of $\Delta V_\text{gap}$ for large $z$ stems from the $\sim 1/z$ behaviour of $V_\text{2B}^+$ limiting $V_\text{2B}^+(z\rightarrow \infty)$ to a finite value, while $V_\text{2B}^-$ linearly increases in absolute value with $z$, \textit{i.e.}, is unrestricted.

\FloatBarrier
\section*{Analytical Transport Model}
Here, the details of the model presented in Fig.~2\textbf{b},\textbf{c} of the main manuscript are presented. The \bb-state is exponentially localized around the \bb-unit, resulting in a double-tunneling-barrier configuration for the tip-\bb-state-substrate system.

The tip-\bb-state separation $d \sim \SI{3.8}{nm}$ is defined by the chemical structure of the ribbon and, thus, is constant. Any change in $z$ translates directly to a change of the \bb-state-substrate separation. Starting from the sketches in Fig.~2\textbf{b} of the main manuscript, resonance conditions for the onset of ballistic transport at either voltage polarity are derived for both ambi- and unipolar transport.

\subsection{Ambipolar Transport}
Resonant transport through the \bb-state is enabled when the associated ion resonance aligns with the Fermi level of tip or substrate. If $V_\text{2B}^-$ corresponds to transport through O2B$^+$ and $V_\text{2B}^+$ to transport through U2B$^-$, respectively, we can directly apply the equations provided by Wagner \textit{et al.} for a molecular quantum dot in scanning quantum dot microscopy \citep{Wagner2015sqdm}
\begin{align}
V_\text{2B}^- &= \frac{\varepsilon_+}{e \alpha} =  \frac{z \varepsilon_+}{e d} \text{ and}\\
V_\text{2B}^+ &= \frac{\varepsilon_-}{e \alpha} = \frac{z \varepsilon_-}{e d},
\end{align}
with the elementary charge $e$ and the two energies $\varepsilon_+$ and $\varepsilon_-$ of the ion resonance associated to the occupied (O2B$^+$) and unoccupied state (U2B$^-$), respectively, and the fraction of voltage $\alpha$ dropping between tip and QD. In general, $\varepsilon_+$ can differ from the energy of the O2B-state ($\varepsilon$), with the same being true for $\varepsilon_-$ and $\varepsilon'$, due to correlation or charging effects. We estimate $\alpha=d/z$ assuming a plate capacitor geometry of tip and substrate, \textit{i.e.} a linear voltage drop over the junction. The blue, dashed line in Figure~2\textbf{c} of the main manuscript presents a fit of the model to the experimental data with $d=\SI{2.6}{nm}$, $\varepsilon_- = \SI{-0.22}{eV}$ and $\varepsilon_+ = \SI{0.5}{eV}$. There is a strong mismatch between the experimental data and the fitted model. For increasing $z$, the model predicts a shift of both $V^-$ and $V^+$ to larger absolute values, increasing the separation between the two states. Our observations clearly contradict this scenario.

\subsection{Unipolar Transport}
Instead, we propose a one-level unipolar transport model to explain the appearance and relative shift of $V_\text{2B}^-$ and $V_\text{2B}^+$, where only the occupied \bb-state is responsible for resonant transport at both voltage polarities (left panel of Fig.~2\textbf{b} of the main manuscript).The onset of resonant transport in this scenario occurs at
\begin{align}
    \varepsilon_+ &= \alpha eV_\text{2B}^-  \text{ and}  \label{eq:neg_unipolar}\\
    -\varepsilon_+ &= (1-\alpha) eV_\text{2B}^+. \label{eq:pos_unipolar}
\end{align}
Solving this set of equations for $V_\text{2B}^-$ and $V_\text{2B}^+$ yields
\begin{align}
    V_\text{2B}^- &= \frac{z \varepsilon_+}{e d} \text{ and} \label{eq:neg_unipolar_solved}\\
    V_\text{2B}^+ &= \frac{- \varepsilon_+ z}{e(z - d)}. \label{eq:pos_unipolar_solved}
\end{align}
Interestingly, the term $(1-\alpha)$ induces a non-linear behaviour in $V_\text{2B}^+$. The unipolar transport model predicts a shift of both $V_\text{2B}^-$ and $V_\text{2B}^+$ towards more negative values in the experimentally relevant range. While $V_\text{2B}^-$ shifts linearly, the $\sim 1/z$ trend of $V_\text{2B}^+$ moves it faster towards negative values at small $z$. This trend levels off at larger $z$. The unipolar model reproduces qualitatively the experimentally observed peak movement. We find empirically, that we need to introduce a constant offset of $V_\text{sd}^+$ on the right side of Equation~(\ref{eq:pos_unipolar_solved}) that translates $V_\text{2B}^+$ to larger voltages. The model incorporating $V_\text{sd}^+$ is fitted to the experimental data and presented in the main manuscript as the red, dashed line in Fig.~2\textbf{c}. We obtain the fit parameters $d=\SI{2.6}{nm}$, $\varepsilon = \SI{-0.22}{eV}$ and $V_\text{sd}^+ = \SI{0.7}{V}$. A possible origin of $V_\text{sd}^+$ is a locally increased voltage barrier for electrons tunneling into the sample caused by the surface dipole moment that builds up as a consequence of the charge transfer between ribbon and substrate \citep{fahlman_interfaces_2019}. This barrier only affects electron tunneling at positive bias.

Unipolar transport through an unoccupied level corresponds to an electron-hole inversion in the model resulting in an exchange of $V_\text{2B}^-$ and $V_\text{2B}^+$, \textit{i.e.}, mirroring their positions at $V=0$. From this consideration, we can exclude unipolar transport through an unoccupied state.

Equation~(\ref{eq:neg_unipolar}) and (\ref{eq:pos_unipolar}) allow us to extract values for $\alpha$ and $\varepsilon_+$ from the experimental $V_\text{2B}^-$ and $V_\text{2B}^+$. The analytical solutions of this set of equations are
\begin{align}
    \alpha &= \frac{V^+}{V^+ - V^-} \text{ and} \label{eq:alpha_unipolar}\\
    \varepsilon_+ &= \frac{eV^- V^+}{V^+ - V^-},  \label{eq:eps_unipolar}
\end{align}
where the index $\text{2B}$ was omitted for clarity. These equations are used in the analysis of the data presented in Fig.~2\textbf{d}, \ref{fig:efig3}\textbf{b},\textbf{c} and \ref{fig:efig4}.

\subsection{Integrating $z$-dependent Fermi-level into the model}
Fig.~2\textbf{d} of the main manuscript hints to a linear Fermi level shift towards lower energy while lifting the ribbon. We introduce such a linear Fermi level shift \textit{a posteriori} in the simulation by adding the linear correction $E_\text{F,0} + zE_\text{F,$z$}$ to Equation~(\ref{eq:neg_unipolar_solved}) and (\ref{eq:pos_unipolar_solved}), where $E_\text{F,0}$ is the (hypothetical) position of the Fermi level for $z=0$ and $E_\text{F,$z$}$ describes its linear shift with $z$. In Fig.~\ref{fig:efig_V2B_z-dependence}, the model including the Fermi level shift (black solid line) is compared with the model presented in the main manuscript (red dashed line). At negative voltage, no difference between the two models is perceivable. At positive voltage, however, the empirical model including the Fermi level shift indeed improves the description of the data slightly. We emphasize though that the non-linear shift of resonance conductance occurring \emph{only} at positive voltage together with the reduction of $\Delta V_\text{gap}$, as we both observe experimentally, excludes a non-linear shift of the Fermi level combined with ambipolar transport as the dominant effect governing the behaviour of the experimental $V_\text{2B}^-$ and $V_\text{2B}^+$.

\section{Analysis and discussion of unipolar transport through the valence band}

\begin{figure}[h!]
    \centering
    \includegraphics[width=1\columnwidth]{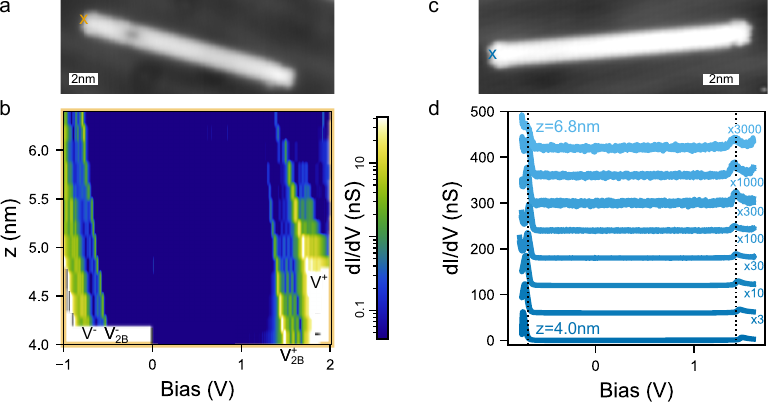}
    \caption{\label{fig:efig1}
    \textbf{Additional Conductance Maps}
    \textbf{a, c} STM images of the \bgnr-2 and the pristine \gnr\ ($V=-\SI{300}{mV}$, $I=\SI{30}{pA}$). The coloured crosses indicate the positions where the ribbons were approached for lifting.
    \textbf{b} Colour map of $\didv$ spectra at different tip sample separation $z$ in logarithmic scale obtained while lifting \bgnr-2.
    \textbf{d} Stack of $\didv$ spectra at different tip sample separation $z$ obtained while lifting the pristine GNR. The spectra are recorded in steps of $\Delta z = \SI{0.4}{nm}$. The spectra are offset by $\SI{55}{nS}$ for clarity. The dashed lines serve as a guide to the eye.
	}
\end{figure}

In the following, the behaviour of resonant band transport \emph{via} $V^-(z)$ and $V^+(z)$ is discussed. Fig.~\ref{fig:efig1}\textbf{a},\textbf{b} presents the STM image and conductance map of a second \bgnr\ (\bgnr-2). This ribbon contains the \bb-dopant at a distance of $\SI{2.9}{nm}$ from the ribbon's termination, \textit{i.e.}, a shorter distance as compared to the value of $\SI{3.8}{nm}$ for the ribbon (\bgnr-1) presented in the main manuscript. For \bgnr-2, two pairs of transport resonances are identified (labelled as $V_\text{2B}^\pm$ and $V^\pm$ in Fig.~\ref{fig:efig1}. Each of them is accompanied by a set of vibronic satellites. The two resonances closest to the Fermi level ($V_\text{2B}^-$ and $V_\text{2B}^+$) are ascribed to transport \textit{via} the O2B$^+$-resonance in agreement with the main manuscript. The second pair of resonances ($V^-$ and $V^+$) is present at elevated voltage and tentatively ascribed to band transport \textit{via} the ribbons valence band (VB), as we will confirm below. Fig.~\ref{fig:efig1}\textbf{c} and \textbf{d} show the STM image of a pristine GNR and a stack of $\didv$-spectra obtained during the lifting of this ribbon. The onset of resonant transport is visible around $V^-\approx -\SI{0.7}{V}$ and $V^+\approx\SI{1.5}{V}$ for the whole data set, despite the drop of conductance by more than three orders of magnitude.

\begin{figure}[t]
	\includegraphics[width=1\columnwidth]{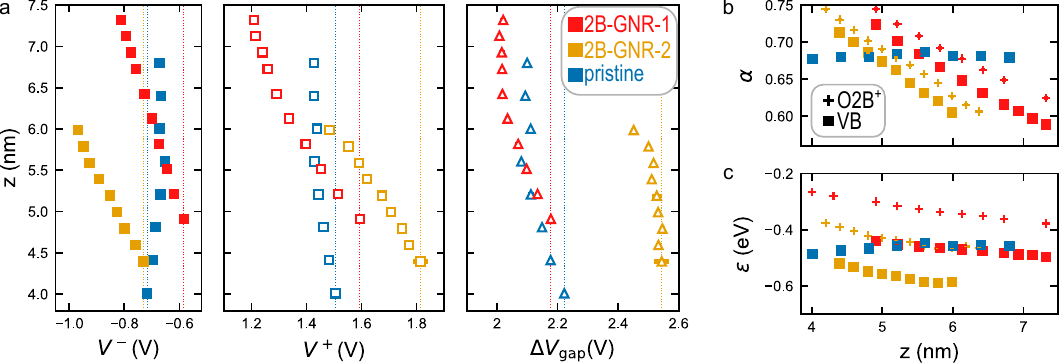}
	\caption{\label{fig:efig3}
	\textbf{Unipolar transport through the valence band:}
	\textbf{a} $V^-$, $V^+$ and $\Delta V_\text{gap}$ as a function of tip-sample separation $z$ for the three presented ribbons. In all cases, the apparent gap $\Delta V_\text{gap}$ reduces during retraction, indicating unipolar transport through the VB. Dotted lines are a guide for the eye.
    \textbf{b} Gating efficiency $\alpha$ as a function of $z$ for transport through the VB (squares) and through O2B$^+$ (crosses). The decrease of $\alpha$ with $z$ observed in the \bb-doped systems indicates their VB behaves as a localized state in a double tunnelling barrier configuration.
	\textbf{c} Energy $\varepsilon$ of the VB (squares) and O2B$^+$ (crosses) as a function of $z$. The significantly lower energy in case of the "orange" \bgnr-2 stems from the confinement of the VB in a smaller segment. The values for transport through O2B (crosses) are presented for easier comparison and discussed in detail in the main manuscript.
	}
\end{figure}

Fig.~\ref{fig:efig3}\textbf{a} presents the position of $V^-$ and $V^+$ as a function of $z$ for the \bgnr-1 presented in the main manuscript (red data points), the \bgnr-2 (orange) and the pristine GNR (blue). In all cases the apparent gap $\Delta V_\text{gap} = V^+ - V^-$ between the band transport resonances closes indicating unipolar transport \textit{via} the VB. Additionally, \bgnr-1 exhibits a \textit{non-linear} shift of $V^+$ in \textit{negative} bias direction, further supporting the unipolar character of resonant electron transport enabled by the VB. Importantly, the closing of the gap is observed for the pristine GNR, too, indicating unipolar transport \textit{via} the ribbon's VB.

A further indication for unipolar resonant band transport is the absolute value of $\Delta V_\text{gap}$ of the pristine and the "red" \bgnr, which are significantly smaller than reported band gaps for ribbons on Au(111) \citep{Wang2016, Carbonell-Sanroma2017}. The hybridization with the metal substrate reduces electron correlations, which results in a reduction of the gap between valence and conduction band in the adsorbed ribbon. If any change were to be expected, it would be a larger band gap.

We employ the same two-barrier model (Equation~(\ref{eq:alpha_unipolar}) and (\ref{eq:eps_unipolar}) ) as before to analyze the unipolar transport through the VB. The results for $\alpha$ are presented in Fig.~\ref{fig:efig3}\textbf{b}. For the pristine ribbon, we obtain $\alpha \approx 0.68$, independent of $z$. This value is in very good agreement with unipolar transport in other molecular wires suspended in an STM junction \citep{Reecht2014}. In fact, a unipolar transport mechanism was proposed recently for modified 7aGNR molecular wires in an independent work \citep{jiang_length-independent_2022}. The constant value  of $\alpha$ indicates that decreasing hybridization with the substrate does not affect the electronic gating of the ribbon.

For the borylated nanoribbons (red and orange squares), $\alpha$ reduces from $\sim 0.7$ to $\sim 0.6$ in the observed region indicating electron transport mediated by a decoupled, localized quantum well state of the VB which arrises due to the presence of the \bb-dopant \citep{Carbonell-Sanroma2017}. The values of $\alpha$ obtained from $V^-$ and $V^+$ are reduced by $\sim 5\%$ with respect to the ones obtained from $V_\text{2B}^-$ and $V_\text{2B}^+$ in the corresponding ribbon (red and orange crosses). Therefore, the hole transport is mediated by the segment of the VB that is confined between tip and \bb-state (referred to as confined VB). This is supported by the different VB energies $\varepsilon_\text{VB}$ discussed below, as well as the quantum well mode presented in Fig.~4\textbf{b} of the main manuscript.

In Fig.~\ref{fig:efig3}\textbf{c}, we show the evolution of the VB energies $\varepsilon_\text{VB}$ (squares) and of the energies of the O2B$^+$ (crosses) with $z$. The pristine ribbon (blue squares) and the the \bgnr-1 (red squares) present a close quantitative agreement between the respective $\varepsilon_\text{VB}$, while the VB of the \bgnr-2 (orange squares) lies at significantly lower energy. In the \bgnr-2, the apex-\bb-unit separation is shorter than in the \bgnr-1, resulting in a ground state energy of the confined VB further away from the Fermi level, as reflected in $\varepsilon_\text{VB}$. The \bb-state remains pinned to the VB with a constant offset of $\sim \SI{150}{meV}$ over the whole observed range for both \bgnr.

\FloatBarrier
\subsection{Rectifying behaviour of the valence band}
\FloatBarrier

\begin{figure}[h!]
    \centering
    \includegraphics[width=1\columnwidth]{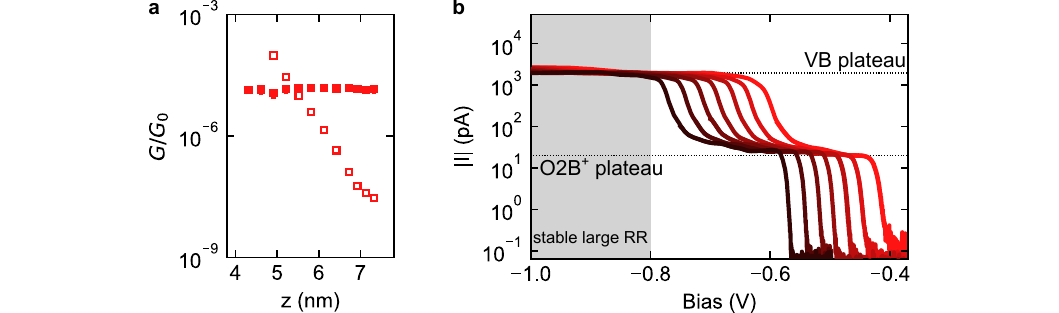}
    \caption{\label{fig:efig-rr-vb}
    \textbf{Current rectification by the valence band}
    \textbf{a} Resonant conductance $G(V^-)$ and $G(V^+)$ as a function of $z$. The same qualitative behaviour is observed as for transport \textit{via} O2B$^+$. We find the decay constants $\beta^- = -0.02\pm\SI{0.02}{nm^{-1}}$ at negative and $\beta^+ = 3.5\pm\SI{0.1}{nm^{-1}}$ at positive bias. 
    \textbf{b} $|I|$-$V$-curves taken from $z=\SI{4.9}{nm}$ (bright red curve) to $\SI{6.7}{nm}$ (dark red curve) in steps of $\Delta z =\SI{0.3}{nm}$. The two different current plateaus associated to the transport opened by the \bb-state and by the VB are clearly distinguished. In the grey shaded area large $\text{RR}>10^5$ are reached for all presented cases.
	}
\end{figure}

The linear conductance $G(V^\pm, z)$ for resonant transport through the VB of the \bgnr-1 follows an identical behaviour as resonant transport through O2B$^+$ (Fig.~\ref{fig:efig-rr-vb}\textbf{a}). The respective decay constants for the VB are $\beta^- = -0.02\pm\SI{0.02}{nm^{-1}}$ at negative and $\beta^+ = 3.5\pm\SI{0.1}{nm^{-1}}$ at positive bias. This conductance characteristics is typical for asymmetric two level systems \citep{Elbing2005, perrin_large_2014, perrin_gate-tunable_2016}. The different onset voltage for resonant conductance at negative and positive voltage allows to maximise the current rectification ratio of the suspended ribbon by choosing $V$ and $z$ such that at negative bias transport through both VB and O2B$^+$ is enabled, while at positive voltage the resonant transport regime is not yet accessible. In this regime, extraordinary large values of $\text{RR}>10^5$ are reached.

Transport through the confined VB is mediated in a similar fashion as transport through O2B$^+$ (compare Fig.~3\textbf{c} of the main manuscript and the discussion there). At negative voltage, the segment of the VB between substrate and \bb-state lies close to the Fermi level, enabling fast hole relaxation to the substrate after the initial tunneling event. The rate-limiting tunneling barrier between tip and \bb-state remains unaffected from changes in $z$, resulting in the constant conductance at resonant transport through the suspended ribbon. However, at inverted voltage polarity, the VB segment between \bb-state and substrate is significantly lower and, therefore, the initial hole tunneling event into the ribbon is exponentially suppressed upon increasing the tip-substrate distance. The hole relaxation into the tip is facilitated by an equivalent mechanism as before.

\FloatBarrier
\section{Vibronic Modes}
\begin{figure}[h!]
    \centering
    \includegraphics[width=0.5\columnwidth]{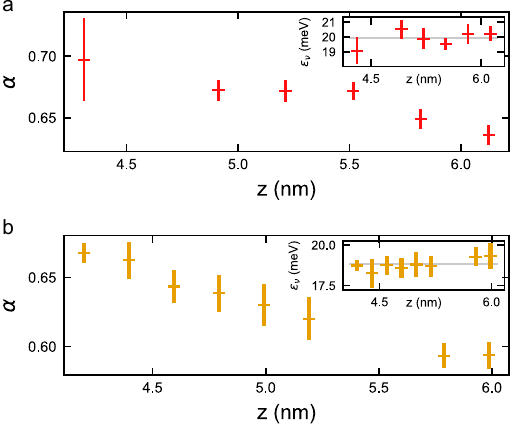}
    \caption{\label{fig:efig4}
    \textbf{Analysis of Vibronic Modes of $V_\text{2B}^\pm$:}
	\textbf{a, b} $\alpha$ for different $z$ for the \bgnr-1 (\textbf{a}) and for the \bgnr-2 (\textbf{b}). The overall trend of decreasing $\alpha$ with increasing $z$ is reproduced, with smaller values for the \bb-unit closer to the tip (\bgnr-2). The inset shows the fundamental vibrational energy remains at a constant value of $E_\nu = 19.9 \pm \SI{0.5}{meV}$ and $E_\nu = 18.8 \pm \SI{0.5}{meV}$, respectively. The $z$-independent character exclude an external vibration of the free-standing segment. The close quantitative agreement between the two measurements supports a relation to a localized vibronic excitation of the boron.
	}
\end{figure}

\FloatBarrier
The vibronic modes present in the \bgnr s undergo the same gating process as the \bb-state and the VB. Therefore, the voltage $V_\nu^\pm$ needed for excitation is different in the two polarities. In order to obtain the fundamental energy of the vibronic mode $\varepsilon_\nu$, we fit a linear regression to the vibronic satellites (inset of Fig.~4\textbf{a} of the main manuscript). The gated excitation voltages $V_\nu^-$ and $V_\nu^+$ obey Eq.~(\ref{eq:alpha_unipolar}) and (\ref{eq:eps_unipolar}). Fig.~\ref{fig:efig4} presents the extracted values of $\alpha$ as a function of $z$ for the \bgnr-1 (\textbf{a}) and \bgnr-2 (\textbf{b}). The reduction of $\alpha$ with increasing $z$ is reproduced clearly. The numerical values are in reasonable quantitative agreement with the $\alpha$ obtained from the O2B$^+$ resonance. 

The fundamental vibronic energy $E_\nu = 19.9 \pm \SI{0.5}{meV}$ (\bgnr-1) and $E_\nu = 19.9 \pm \SI{0.5}{meV}$ (\bgnr-2), presented in the insets of the respective panels, does not correspond to internal vibronic modes of a pristine 7aGNR \citep{Overbeck2019} or of a 7aGNR with dense \bb-doping \citep{Senkovskiy2018}. The $z$-independent character of $E_\nu$ excludes an external vibration of the free-standing segment of the ribbon. The energy of a longitudinal compressive mode \cite{Overbeck2019} of the pristine GNR segment between tip and \bb-state does not match our $E_\nu$ by a factor of 2 and is expected to differ significantly between the two ribbons.

\begin{figure}[h!]
    \centering
    \includegraphics[width=1\columnwidth]{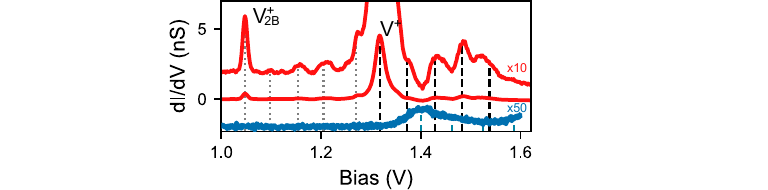}
    \caption{\label{fig:efig2}
    \textbf{Comparison of Vibronic Modes of $V_\text{2B}^+$ and $V^+$}
    $\didv$ spectrum of the \bgnr-1 taken at $z=\SI{6.1}{nm}$ (red) and the pristine GNR at $z=\SI{6.0}{nm}$ (blue). The energy spacing between the satellite peaks above $V_\text{2B}^+ = \SI{1.047}{V}$ and $V^+ = \SI{1.318}{V}$ is identical, indicating that they correspond to the same phonon mode. For the pristine ribbon the expected voltages associated to that phonon mode are indicated, but resonances are absent in this case.
	}
\end{figure}

Interestingly, we observe the vibronic modes also as satellite peaks to the VB in case of the \bgnr. Fig.~\ref{fig:efig2} presents the $didv$-spectrum of \bgnr-1 at $z=\SI{6.1}{nm}$ (red curves). The vibronic resonances associated to $V_\text{2B}^+$ are indicated as dotted lines. The dashed lines indicate expectation values for vibronic excitations above the VB $V_{\nu\text{,VB}} = V^+ + n \alpha E_\nu /e$, where $n = 1, 2, 3, 4$ and $\alpha=0.65$ is taken from Fig.~\ref{fig:efig3}\textbf{b}. A set of resonances perfectly coincides with the predicted vibronic modes. On the other hand, in the pristine GNR (blue curve) no satellite resonances are visible at equivalently calculated voltages. The absence of the vibronic modes in the pristine ribbon clearly indicates the vibronic mode is associated to the \bb-dopant.

\FloatBarrier
\section{Particle-in-a-Box Model}
The energy of a particle in an infinite quantum well is 
\begin{align}
    \varepsilon_n = \frac{\pi^2 \hbar^2}{2 m L^2} n^2,
\end{align}
with the reduced Planck constant $\hbar$, the particle mass $m$, the length of the quantum well $L$ and the state number $n \in \mathbb{N}$. We assume a confinement length of $L=\SI{3.8}{nm}$, corresponding to the geometrical distance between ribbon termination and \bb-unit, and the effective electron mass of $m_\text{eff} = 0.4 m_\text{e}$ \citep{Sode2015} to calculate the energy difference between the first two quantum well modes. The energy is converted to applied bias by considering the experimental values of $\alpha (z)$ at negative and $(1-\alpha (z))$ at positive bias, respectively (Figure~4b of the main manuscript).

\FloatBarrier